\documentclass[aps,prx,floatfix,reprint,twocolumn,showpacs,preprintnumbers,superscriptaddress,amsmath,amssymb]{revtex4-1}

\usepackage{hyperref}
\hypersetup{colorlinks=true, citecolor=blue, urlcolor=blue, linkcolor=blue}
\usepackage{graphicx}  
\usepackage{dcolumn}   
\usepackage{bm}        
\usepackage{stmaryrd}  
\usepackage{multirow}  
\usepackage{color}     
\usepackage{amsmath,amssymb}
\usepackage{enumerate}
\usepackage{float}
\usepackage{mathtools}
\usepackage[dvipsnames]{xcolor}

\newcommand{\pder}[2]{\frac{\partial #1}{\partial #2}}            

\begin{document}

\preprint{APS/123-QED}

\title{Interaction of in-plane magnetic skyrmions with $90^\circ$ magnetic domain walls: micromagnetic simulations}

\author{Pavel Bal\'a\v{z}} 
\email{balaz@fzu.cz}
\affiliation{FZU -- Institute of Physics of the Czech Academy of Sciences, Na Slovance 1999/2, 182 21 Prague 8, Czech Republic}

\date{\today}

\begin{abstract}
$90^\circ$ pinned magnetic domain walls can be observed in thin magnetic layers attached to a ferroelectric substrate.
The main stabilization mechanism of the noncollinear magnetic texture is the strain transfer which is responsible
for imprinting of the ferroelectic domains into the uniaxial anisotropy of the ferromagnet.
Here, we investigate by means of micromagnetic simulations 
how the interfacial Dzyaloshinkii-Moriya interaction influences the $90^\circ$ domain wall 
structure.
It is shown that Dzyaloshinskii-Moriya interaction induces a large out-of-plane magnetization component, 
strongly dependent on the domain wall type.
In particular, it is shown that this out-of-plane magnetization component is crucial for the transport
of the in-plane magnetic skyrmions, also known as bimerons, through the magnetic domain walls.
Based on the results of micromagnetic simulations, a concept of in-plane magnetic skyrmion valve based
on two $90^\circ$ pinned magnetic domain walls is introduced.
\end{abstract}


\maketitle

\section{Introduction}
\label{Sec:Intro}

Coupling of ferroelectric and ferromagnetic degrees of freedom is the key ingredient of novel multiferroic
structures, which are intensively studied for their considerable potential 
for the future applications in spintronics~\cite{Gradauskaite:PSR_2021}.
It has been demonstrated that when a thin ferromagnetic layer is attached to a ferroelectric one,
the ferroelectric domain patterns might imprint into the structure of the uniaxial anisotropy in the
ferromagnet. As a result, magnetic anisotropy axis abruptly rotates at the anisotropy boundary by
$90^\circ$ giving rise to the $90^\circ$ magnetic domain walls (DWs) strongly pinned to the 
position of the anisotropy boundary~\cite{Franke:PRL_2014}.
The physical mechanisms recognized beyond this effect are exchange 
coupling between the canted magnetic moment of ferroelectric layer and an adjacent ferromagnetic
one~\cite{Lebeugle:PRL_2009,Heron:PRL_2011} 
and strain transfer from ferroelectric domains to a magnetostrictive 
film~\cite{Lahtinen:IEEETransMagn_2011,Lahtinen:AdvMat_2011,Lahtinen:APL_2012,Chopdekar:PRB_2012}.
According to the in-plane magnetic texture, 
two types of the $90^\circ$ magnetic DWs can be distinguished in the ferroelectric/ferromagnet bilayers; 
uncharged and charged. Both of them are strictly in-plane magnetic textures.
In the center of an uncharged $90^\circ$ magnetic DW, the magnetization is perpendicular to the magnetic DW.
In contrast, magnetization in the center of a charged $90^\circ$ DW is aligned with the DW.
While the first type is stabilized mainly by the interplay of the magnetic exchange interactions 
with uniaxial anisotropy, the latter one is usually influenced by the magnetostatic dipolar interaction~\cite{Franke:PRL_2014}.

The technical merit of the $90^\circ$ magnetic DWs in multiferroic heterostructures 
consists in a vast array of related physical mechanisms that can be deployed in future applications.
Firstly, it has been demonstrated that the anisotropy boundaries in the ferromagnet dynamically follow 
the changes in the ferroelectric domain structure~\cite{FRA-15}. 
This effect allows one to position the ferromagnetic DWs purely via the electric field.
Secondly, a possibility of utilizing a $90^\circ$ pinned DW as a highly efficient source 
of spin waves in a narrow range of
wave lengths has been shown by means of micromagnetic simulations~\cite{VanDeWiele:SciRep_2016}.
Moreover, a theoretical analysis has shown that the static and dynamic properties of the
pinned $90^\circ$ DWs can be tuned via applied magnetic field~\cite{Balaz2018:PRB}.
Alternatively, short-wavelength spin waves can be emitted in multifferoic heterostructures 
with modulated magnetic anisotropy using applied magnetic field~\cite{HAM-17}.
Finally, it has been recently reported that individual magnetic domains 
in multiferroic composite bilayers can be manipulated via a laser pulse showing laser-induced 
changes of magnetoelastic anisotropy leading to precessional switching~\cite{Shelukhin:PRAppl_2020}.

In this paper, we extend the range of phenomena discussed in the anisotropy-modulated multiferroic multilayers
by examining the effect of interfacial Dzyaloshinskii-Moriya 
interaction (DMI)~\cite{Dzyaloshinsky_1958,Moriya_1960} on the $90^\circ$ pinned DWs 
employing micromagnetic simulations.
DMI can be observed not only in bulk crystals of low symmetry~\cite{Dzyaloshinsky_1958} 
but also in layered materials where the inversion symmetry is broken at the
interfaces~\cite{Crepieux:JMMM_1998,Hrabec:PRB_2014,Je:PRB_2013,Cho:NatComm_2015,Kim:PRB_2018}.
It has been shown that DMI significantly affects the dynamics of $180^\circ$ magnetic DWs
by increasing their Walker field and velocity~\cite{Thiaville:EPL_2012,Garcia:PRB_2021}. 
DMI has also a noticeable impact on the critical current density for DW depinning 
by spin transfer torque and DW resonance frequency~\cite{Li:JAP_2015}.
Here, we show that interfacial DMI might substantially influence
the structure of the $90^\circ$ DWs by inducing an out-of-plane magnetization component.
While this effect introduces just a minor variation of the magnetization texture of the uncharged DWs,
it becomes conspicuous in the case of the charged DWs.
Importantly, the direction of the out-of-plane magnetization induced by DMI
depends on the direction of the in-plane magnetization rotation.
Thus the out-of-plane magnetization components are opposite for the head-to-head and tail-to-tail charged DWs.

Recently, numerical simulations by Moon {\em et al.} have shown a possibility of existence of
the in-plane skyrmions in thin magnetic layers with in-plane easy-axis 
magnetic anisotropy and interfacial DMI~\cite{Moon:PRAppl_2019}.
Similar topological magnetic structures in systems with in-plane magnetic anisotropy have been
reported in various materials and models~\cite{Zhang:SciRep_2015,Kharkov:PRL_2017,Gobel:PRB_2019,Li:NPJCM_2020}.
Existence of the in-plane skyrmions would open new
possibilities to study their interactions with other planar magnetic textures like the in-plane magnetic DWs.
Another advantage of the in-plane skyrmions is that, 
unlike skyrmions in systems with the perpendicular magnetic anisotropy, 
the in-plane skyrmions of both topological charges ($Q = \pm 1$) can simultaneously exist in the same magnetic domain. 
Importantly, due to the skyrmion Hall effect~\cite{Jiang:Nature_2017,Litzius:Nature_2017}, 
current-induced trajectories of skyrmions of opposite charges
are bent in opposite directions~\cite{Moon:PRAppl_2019}.
Although, the mentioned skyrmions are in-plane, their core magnetization is perpendicular to the layer.
We show here that DMI-induced out-of-plane magnetization in the charged DWs can significantly influence
the interaction of the in-plane skyrmions with DWs.
Namely, using micromagnetic simulations we demonstrate that charged pinned $90^\circ$ magnetic DWs can serve
as a topological charge selective skyrmion filters.
Following the results of micromagnetic simulations, we introduce a simple concept of 
in-plane skyrmion valve based on two $90^\circ$ pinned magnetic DWs.

The paper is organized as follows. In Sec.~\ref{Sec:DWs} we describe studied systems of $90^\circ$ magnetic DWs
in a layer with modulated easy-axis and our results of micromagnetic simulations.
Section~\ref{Sec:Skyrmions} analyzes in-plane skyrmions and their dynamics in our system.
Consequently, in Section~\ref{Sec:SkyrmionsDW} we describe mutual interaction of in-plane magnetic skyrmions with the $90^\circ$ magnetic DWs. 
In Sec.~\ref{Sec:Valve} we describe the concept of in-plane skyrmion valve
and discuss its functioning in practice.
Finally, we conclude in Sec.~\ref{Sec:Discussion}. 
Additional materials can be find in the Appendix.

\section{$90^\circ$ magnetic domain walls in the presence of DMI}
\label{Sec:DWs}

\begin{figure}[htp]
\centering
	\includegraphics[width=.99\columnwidth]{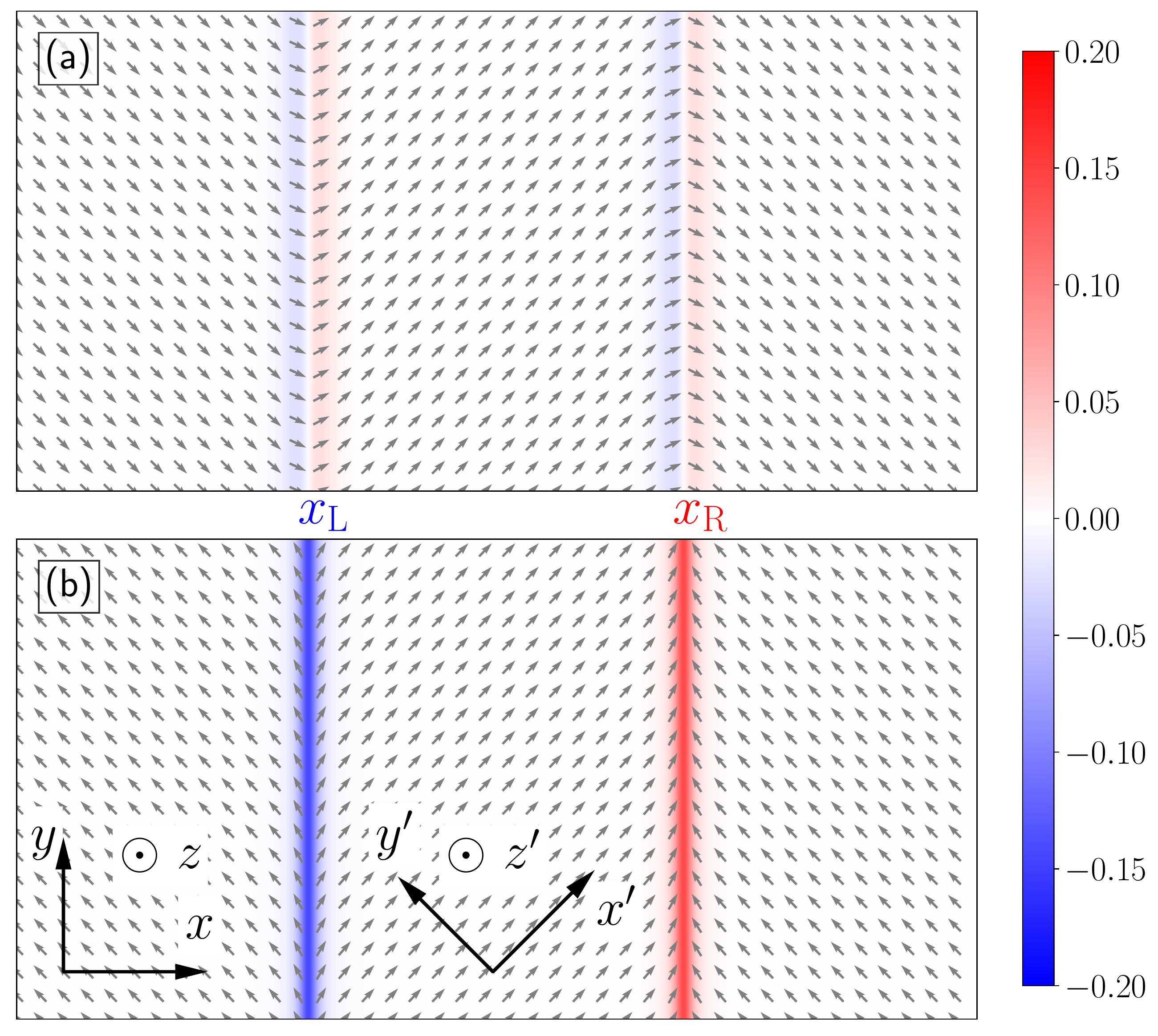}
	\caption{\label{Fig:dw90deg_dmi} 
	$90^\circ$ magnetic DWs resulting from micromagnetic simulations: (a) uncharged DWs, (b) charged DWs.
	The color scale shows the $z$-component of reduced magnetization vector ${\bm m} = {\bm M}/M_{\rm s}$.
	The parameters of calculation were $A_{\rm ex} = 1.3 \times 10^{-11}\, {\rm J}/{\rm m}$, 
	$M_{\rm s} = 7.20 \times 10^5\, {\rm A}/{\rm m}$, $K_u = 10^{4}\, {\rm J}/{\rm m}^3$, and 
	$D = 2\, {\rm mJ}/{\rm m}^2$. The distance between the domain walls is 
	$x_{\rm R} - x_{\rm L} = 1\, \mu m$.}
\end{figure}

First, we shall analyze, how DMI influences the $90^\circ$ magnetic DWs pinned to the anisotropy boundaries.
To this end, we employ micromagnetic simulations implemented in the MuMax3 framework~\cite{mumax3}.
We simulated a thin magnetic layer of thickness $d = 1\, {\rm nm}$, shown in Fig.~\ref{Fig:dw90deg_dmi}.
The lateral sized of the layer were $L_x = 1024\, \Delta x$ and $L_y = 512\, \Delta y$, where
$\Delta x = \Delta y = 2.5\, {\rm nm}$ are the sizes of the discretization cell 
along the $x$ and $y$ axis, respectively.
The discretization cell size along the $z$-axis is equal to the layer's thickness, $\Delta z = d = 1\, {\rm nm}$.

In the layer's plane we defined two anisotropy boundaries, where easy axis abruptly changes its direction
by $90^\circ$. The anisotropy boundaries are located at positions $x = x_{\rm L}$ and $x_{\rm R}$,
shown in Fig.~\ref{Fig:dw90deg_dmi}. 
The distance between the anisotropy boundaries was set to $|x_{\rm R} - x_{\rm L}| = 1\, \mu{\rm m}$.
In the simulations we assumed periodic boundary conditions along the $x$ and $y$ axis.
Therefore, the studied system is translationally symmetric along the $y$-axis.
The easy axis changes along the $x$ direction as follows
\begin{equation}
	\hat{\bm e}_u = 
	\begin{cases}
		\left(1 / \sqrt{2}, 1 / \sqrt{2}, 0 \right) & \text{for} \quad x \in (x_{\rm L}, x_{\rm R})\\
		\left(1 / \sqrt{2}, -1 / \sqrt{2}, 0 \right) & \text{otherwise}\,.
	\end{cases}
\end{equation}
The uniaxial magnetic anisotropy parameter, $K_u$, is the same in the whole layer.
Moreover, in the simulations we used following parameters:
exchange stiffness parameter $A_{\rm ex} = 1.3 \times 10^{-11}\, {\rm J}/{\rm m}$,
saturated magnetization $M_{\rm s} = 7.20 \times 10^5\, {\rm A}/{\rm m}$~\cite{Moon:PRAppl_2019},
corresponding to the range of parameters of cobalt and iron based 
magnetic thin films~\cite{Piao:APL_2011,Chaves:JAP_2015,Yamanouchi:IEEEML_2011,Liu:AIPAdv_2016}.

Furthermore, we assumed interfacial DMI, as described in Refs.~\cite{mumax3} and \cite{Bogdanov:PRL_2001},
given by an effective field term
\begin{equation}
	{\bm H}_{\rm DM} = \frac{2\, D}{\mu_0\, M_{\rm s}} \left(
		\pder{m_z}{x}, \pder{m_z}{y}, -\pder{m_x}{x} -\pder{m_y}{y}
	\right)\,,
\label{Eq:HDM}
\end{equation}
where $\mu_0$ is the vacuum permeability,
${\bm m} = (m_x, m_y, m_z)$ is a unit vector in the direction of local magnetization, 
${\bm M} = M_{\rm s} {\bm m}$, and $D$ is the strength of the DMI.
In our simulations with nonzero DMI, we assumed 
$D = 2\, {\rm mJ}/{\rm m}^2$~\cite{Moon:PRAppl_2019, Yang:PRL_2015}.

Figure~\ref{Fig:dw90deg_dmi} shows stable $90^\circ$ DWs, calculated using micromagnetic simulations,
under the effect of the DMI. 
The anisotropy parameter used in the simulations was $K_u = 10^{4}\, {\rm J}/{\rm m}^3$.
The in-plane magnetization components are shown by the arrows, while the color scale
expresses the out-of-plane, $m_z$, component.
Fig.~\ref{Fig:dw90deg_dmi}(a) depicts the uncharged DWs. 
Charged DWs can be seen in Fig.~\ref{Fig:dw90deg_dmi}(b). 
In both cases we can see out-of-plane magnetization localized at the DWs.
In case of the uncharged DWs the out-of-plane magnetization is rather small and changes sign when passing the
anisotropy boundary.
On the other hand, out-of-plane magnetization of the charged DW is about 10 times larger than the one
of the uncharged DW. 
Moreover, we can see that the out-of-plane magnetization
of the tail-to-tail DW, located at $x_{\rm L}$, has a sign opposite to the one of the head-to-head DW, 
at $x_{\rm R}$.

\begin{figure}
\centering
	\includegraphics[width=.99\columnwidth]{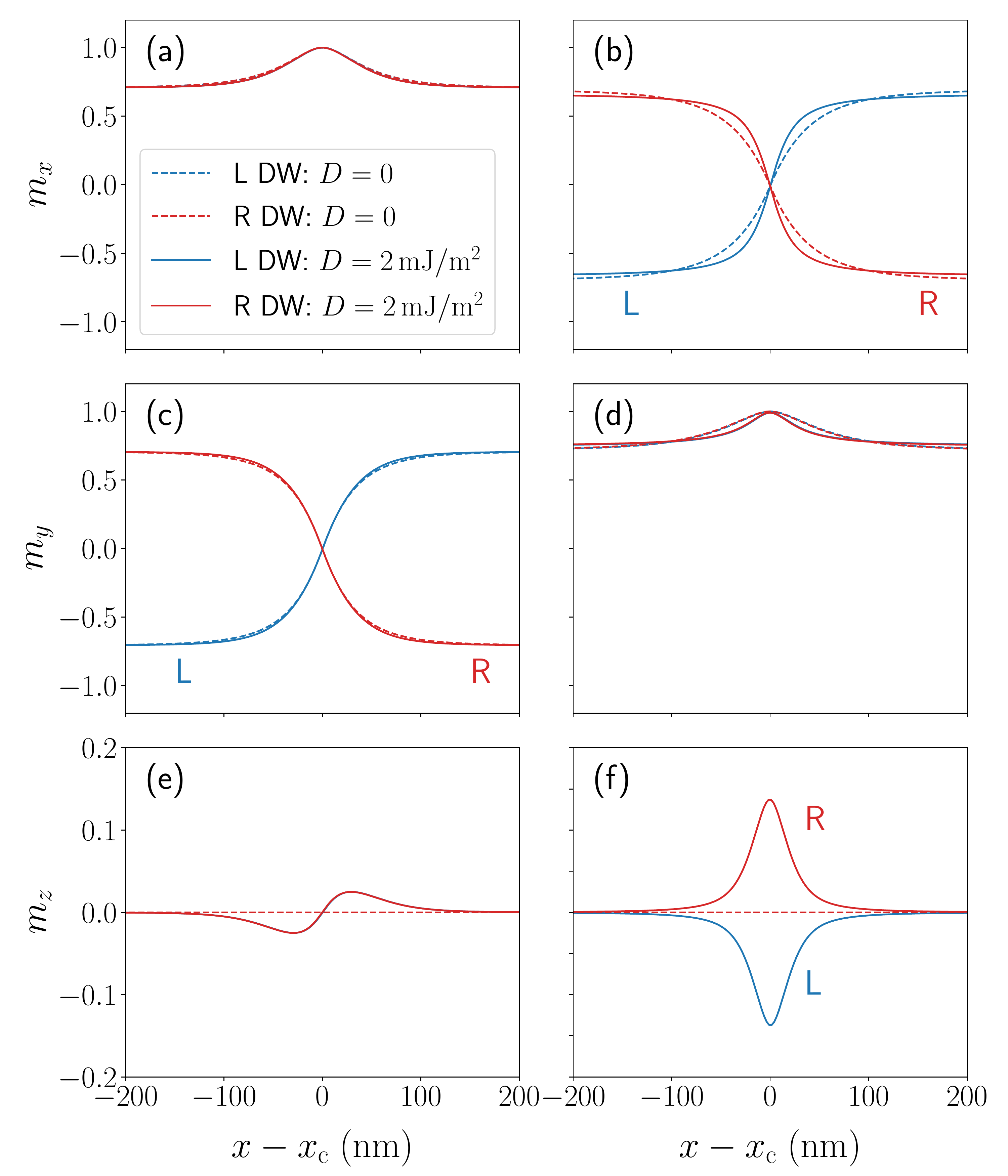}
	\caption{ \label{Fig:DW90deg_prof}
	Comparison of magnetization profiles at the anisotropy boundaries without (dashed lines) 
	and with (solid lines) DMI.
	Magnetization components of the uncharged DWs are shown in the left column, while the ones
	of the charged DWs are plotted in the right one.	
	Components $m_x$, $m_y$, and $m_z$ are plotted in the panels (a)--(b), (c)--(d), and (e)--(f), respectively.
	The parameters of the calculations 	are the same as in Fig.~\ref{Fig:dw90deg_dmi}.
	The magnetization components are plotted as a function of position $x$ with respect to the
	anisotropy boundary, $x_{\rm c}$; $x_{\rm c} = x_{\rm L}$ for the left DW (L), and 
	$x_{\rm c} = x_{\rm R}$ for the right (R) DW.}
\end{figure}
To understand the effect of the DMI, let us compare the domain wall profiles, plotted
in Fig.~\ref{Fig:DW90deg_prof}, showing the magnetization components
at the anisotropy boundaries calculated without and with the DMI.
The magnetization components are plotted as a function of position $x$ (at constant $y$) 
with respect to the anisotropy boundary position, $x_{\rm c}$; 
$x_{\rm c} = x_{\rm L}$ for the left DW, and 
$x_{\rm c} = x_{\rm R}$ for the right DW.
Magnetization components calculated without and with the DMI are plotted by the dashed and solid lines, 
respectively.
Note, the local magnetization vector is normalized to 1.
The in-plane magnetization components of the the uncharged and charged DWs are shown in
Fig.~\ref{Fig:DW90deg_prof}(a) -- (d).
The left column shows magnetization components of the uncharged DWs,
while the right one plots the magnetization components of the charged DWs.
First, the uncharged DWs seem to be almost uninfluenced by the DMI.
The charged DW, however, appears to be significantly affected by the DMI field, which
reduces the DW width.
Similar trends can be observed also on the DMI effect on the out-of-plane magnetization components
plotted in Fig.~\ref{Fig:DW90deg_prof}(e) and Fig.~\ref{Fig:DW90deg_prof}(f).
In case of $D = 0$ the uncharged as well as charged DWs are
strictly planar structures with zero out-of-plane magnetization components~\cite{Balaz2018:PRB}.
Thus, the out-of-plane magnetization observed in Fig.~\ref{Fig:dw90deg_dmi} are induced 
purely by the DMI. This effect is understandable, since the Dzyaloshinskii-Moriya field,
${\bm H}_{\rm DM}$, has a non-zero $z$-component, which is proportional to $\partial m_x/ \partial x$.
Although, DMI induces just a minor out-of-plane magnetization component in the uncharged DWs,
in case of the charged DWs the effect of DMI is about $10$-times stronger.
The reason, why the uncharged DWs are almost unaffected by the DMI is
that it is defined mainly by variation of the $m_y$ component
with modest change of $m_x$.
On the other hand, in case of charged DWs, dominant magnetization component which changes along the $x$ direction
is $m_x$.
Additionally, the charged DWs have also relatively large magnetic moment which responses to the
Dzyaloshinskii-Moriya field by tilting in the $z$-direction.
The direction of the DMI-induced out-of-plane magnetization of the charged DWs 
can be explained by Eq.~(\ref{Eq:HDM}), since $\partial m_x/ \partial x$
has a different sign in case of the tail-to-tail and head-to-head DW.
In contrast, the out-of-plane components of the right and left uncharged DWs are both the same
since the $m_x$ components vary in the same way in both cases.

Our simulations confirm that a charged magnetic DW has higher energy than the uncharged one.
The difference in energy densities of the two systems shown in Figs.~\ref{Fig:dw90deg_dmi} (a) and (b) is
$\Delta\varepsilon \sim 124\, {\rm mJ}/{\rm m}^2$. As the magnetic anisotropy increases, 
total energy of both systems decreases linearly keeping approximately the same energy difference 
between the two systems. The main reason of elevated energy of the charged $90^\circ$ magnetic DW
is its relatively large magnetostatic field.

In our micromagnetic simulations, the distance between the DWs was set to $1\, \mu{\rm m}$.
In order to rule out that the dipolar interaction between the two DWs influences the 
final magnetic configuration, we simulated longer sample with $L_x = 2048\, \Delta x$
with the other parameters unchanged, where distance between the DWs has been
set to $2\, \mu{\rm m}$. The final magnetic configurations matched with those presented
in Figs.~\ref{Fig:dw90deg_dmi} and \ref{Fig:DW90deg_prof}.
Thus, we can exclude any effect of the dipolar coupling between the charged DWs 
on their magnetic configurations.

\subsection{Effect of magnetic anisotropy}

\begin{figure}[htp]
\centering
	\includegraphics[width=.8\columnwidth]{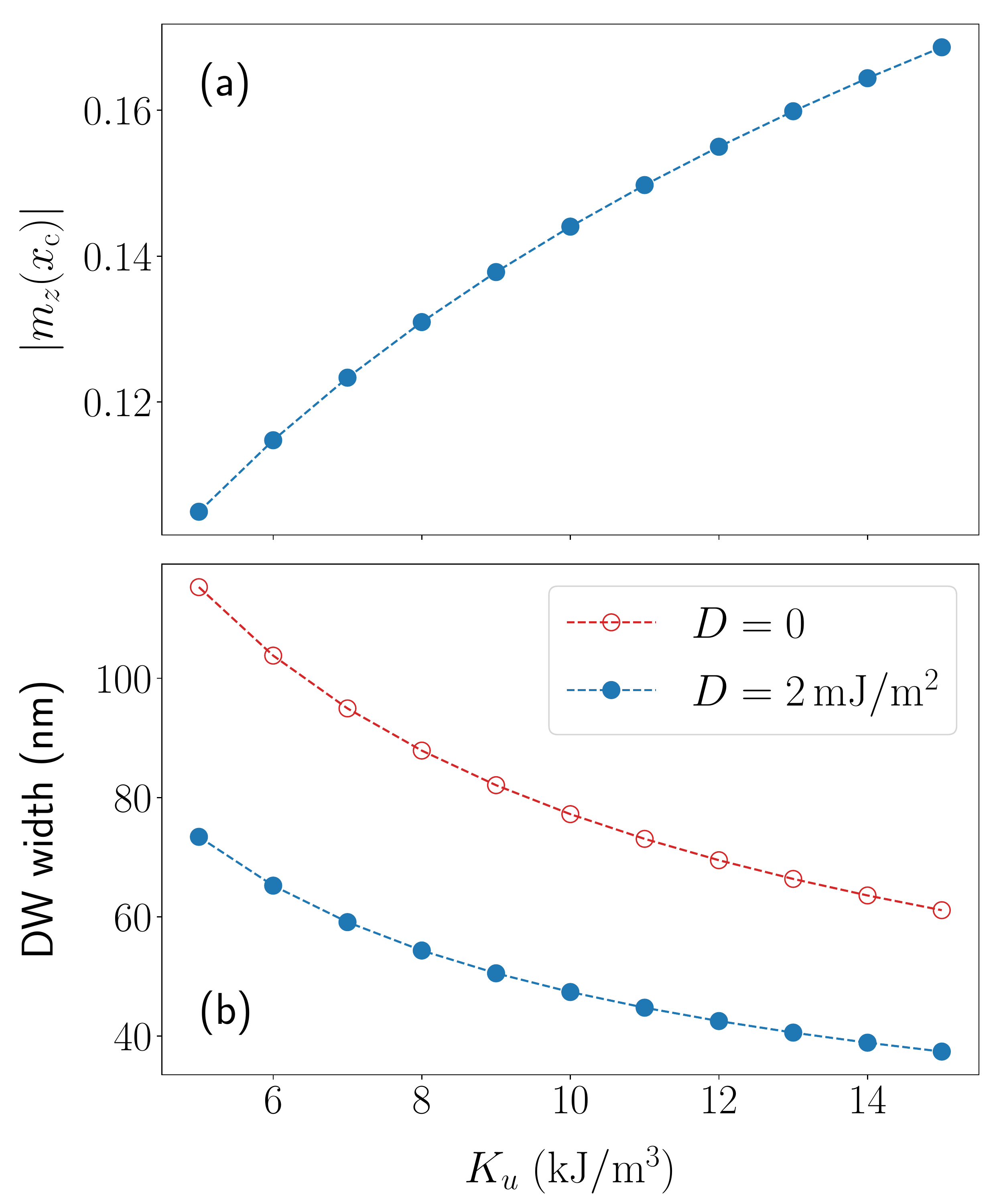}
	\caption{ \label{Fig:Ku_dep}
		Dependence of (a) the out-of-plane magnetization, $m_z(x_{\rm c})$, and 
		(b) the DW width of a charged $90^\circ$ DW as a function of the anisotropy constant, $K_u$.
		The parameters used in the calculations are the same as in Fig.~\ref{Fig:dw90deg_dmi}.
		Calculations of $m_z(x_{\rm c})$ were done with $D = 2\, {\rm mJ}/{\rm m}^2$;
		$x_{c}$ is the position of the anisotropy boundary.
		DW width has been calculated with and without DMI.}
\end{figure}
Let us now inspect how the charged DW structure depends on the amplitude of the uniaxial anisotropy, $K_u$.
Fig.~\ref{Fig:Ku_dep}(a) shows how the DMI-induced out-of-plane magnetization changes as a function of $K_u$.
Fig.~\ref{Fig:Ku_dep}(b) compares DW width with (solid circles) and without (open circles) DMI.
The DW width has been determined from the results of micromagnetic simulations as~\cite{Franke:PRL_2014}
\begin{equation}
	\delta = \int_{-\infty}^{\infty} {\rm d}x\; \cos^2 \left( \phi'(x) \right)\,,
\label{Eq:DW_width}
\end{equation}
where
\begin{equation}
	\phi'(x) = \left[ \phi(x) - \frac{1}{2}\Delta\phi \right] \frac{\pi}{\Delta\phi}
\end{equation}
with $\phi(x)$ being the magnetization angle calculated from the easy axis in the central domain of the
simulated sample. Moreover, $\Delta\phi$ is the magnetization rotation angle calculated as a difference of
magnetization angles in the left and right domains far from the domain wall, 
$\Delta\phi = \phi_{\rm R} - \phi_{\rm L}$.

First, the out-of-plane magnetization induced by DMI monotonously increases with $K_u$.
This effect can be elucidated simply, by shrinking of the DW width as $K_u$ rises,
as shown in Fig.~\ref{Fig:Ku_dep}(b).
Since the DW width decreases, the derivation $\partial m_x / \partial x$ at the anisotropy boundary increases
and, consequently, the $z$-coordinate of the Dzyaloshinkii-Moriya field, ${\bm H}_{\rm DM}$, becomes higher.
Second, Fig.~\ref{Fig:Ku_dep}(b) demonstrates that the DW width decreases with $K_u$ for zero as well as for nonzero $D$.
For the case with $D = 0$, width of a 90$^\circ$ charged DW can be approximated 
as $\delta \simeq \pi \sqrt{A / (2 K_u)}$. 
Introducing DMI leads to a significant reduction of DW width, which is caused by the out-of-plane magnetization component.

\subsection{Effect of magnetic field}

\begin{figure}
\centering
	\includegraphics[width=.95\columnwidth]{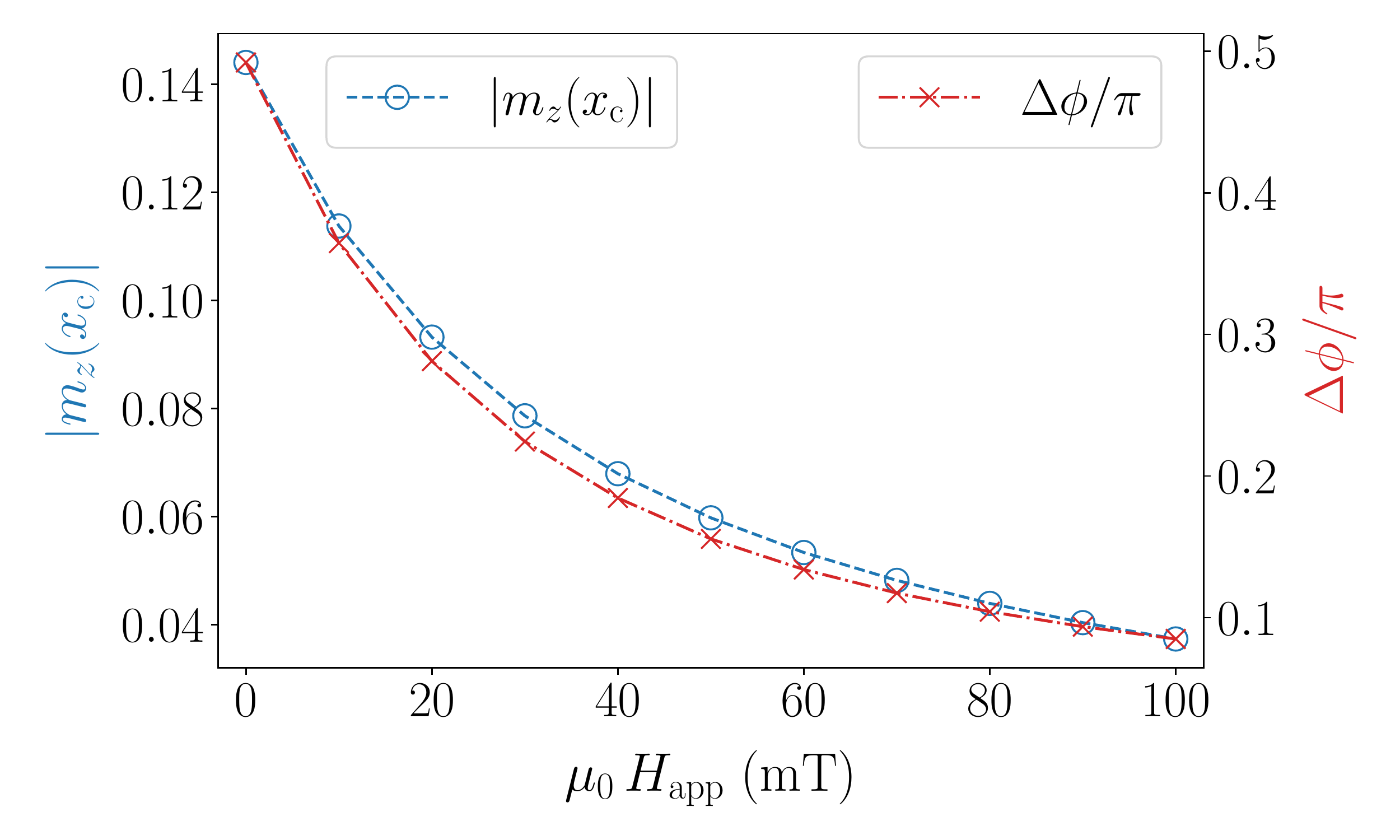}
	\caption{\label{Fig:B_dep}
		Dependence of the DMI-induced magnetization as a function of the magnetic field applied along the $y$-axis. 
		The equilibrium magnetization rotation angle $\Delta\phi$ is plotted on the right vertical axis.}
\end{figure}
Next, we analyze how the out-of-plane magnetization induced by DMI depends on the applied magnetic field.
When the magnetic field is applied along the $y$ direction, ${\bm H}_{\rm app} = H_{\rm app} \hat{\bm e}_y$,
with $H_{\rm app} > 0$, the equilibrium magnetization rotation angle decreases. 
Figure~\ref{Fig:B_dep} plots field dependence of the out-of-plane magnetization component 
of a charged DW at the anisotropy boundary.
When magnetic field, is applied along the DW, the magnetization rotation angle decreases~\cite{Balaz2018:PRB}, 
as shown on the right vertical axis. 
This results in the reduction of the out-of-plane magnetization at the anisotropy boundary.
The effect is the same for both head-to-head and tail-to-tail DWs. 
As $\Delta\phi$ vanishes, $m_z(x_{\rm c})$ also tends to zero.

\subsection{Effect of the spin transfer torque}
\label{SSec:current}

\begin{figure}
\centering
	\includegraphics[width=.8\columnwidth]{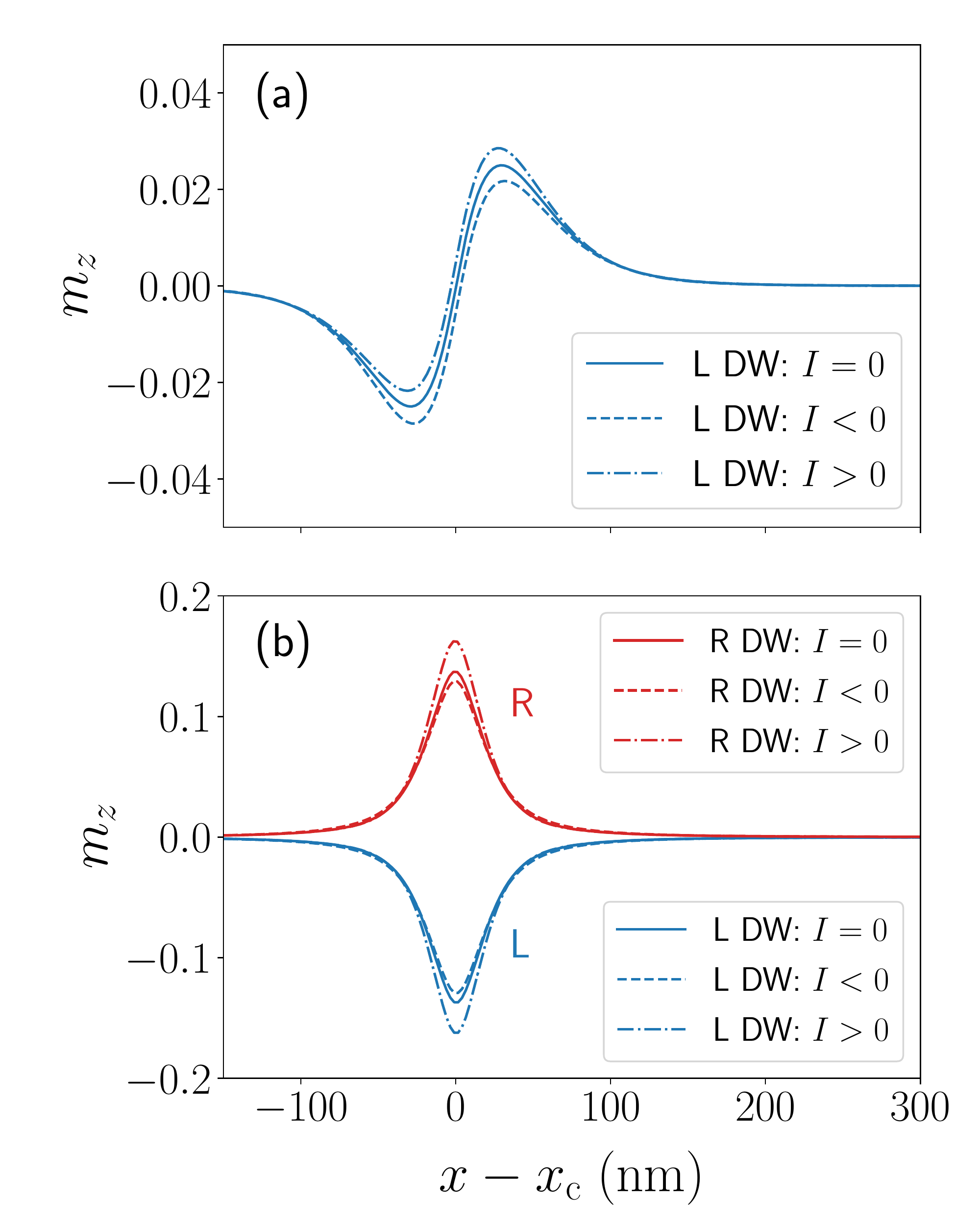}
	\caption{\label{Fig:current}
	Effect of the spin transfer torque on the out-of-plane DW profile
	in case of a (a) uncharged and (b) charged $90^\circ$ DW.
	The anisotropy boundary is located as $x_{\rm c}$ 
	($x_{\rm c} = x_{\rm L}$ for the left (L) DW and $x_{\rm c} = x_{\rm R}$ for the right (R) DW).
	In all the calculations we assumed $D = 2\, {\rm mJ}/{\rm m}^2$.
	The electric current flows in the direction of the $x$-axis. 
	When the current density, $I$, is non-zero, it has 	been set to $I = \pm 10^{12}\, {\rm A}/{\rm m}^2$.
	The other simulation parameters are the same as in Fig.~\ref{Fig:dw90deg_dmi}.
	In case of the uncharged DWs, the current-induced effect for the left and right DW is identical.}
\end{figure}

$90^\circ$ magnetic DW can be excited by electric current flowing in the magnetic layer~\cite{VanDeWiele:SciRep_2016}.
It has been also shown by means of numerical simulations that electric current induces
a small out-of-plane magnetization in the vicinity of the anisotropy boundary
due to the spin transfer torque~\cite{Balaz2018:PRB}.
In order to compare the current-induced and DMI-induced out-of-plane magnetization components, we
carried out number of simulations with DMI with electric current applied in the direction perpendicular to DW.
The effect of the spin transfer torque is simulated using the Li-Zhang term~\cite{Li:PRL_2004,Li:PRB_2004} 
as implemented in MuMax3~\cite{mumax3} taking into account the Gilbert damping parameter
$\alpha = 0.3$ and the spin torque nonadiabaticity $\beta=0.6$.
The results are shown in Fig.~\ref{Fig:current}, which compares the out-of-plane magnetization profile for   
uncharged and charged DWs with and without electric current flowing along the $x$ direction.
When the current density, $I$, was non-zero, we set $I = \pm 10^{12}\, {\rm A}/{\rm m}^2$,
which is a typical value used for current-induced DW dynamics.
In all cases we assumed $D = 2\, {\rm mJ}/{\rm m}^2$, and $K_u = 10^{4}\, {J}/{\rm m}^3$.

Spin transfer torque induces changes of the out-of-plane magnetization in case of 
the uncharged as well as charged DW. This change is, however, about one order of magnitude smaller 
than the one induced by DMI.
In case of the charged DW, out-of-plane magnetization component increases when $I>0$, 
however, it reduces for opposite current direction.
Nevertheless, in the studied system, electric current has just a modest effect on the magnetization profile.

\section{In-plane skyrmions}
\label{Sec:Skyrmions}

\begin{figure}
\centering
	\includegraphics[width=.75\columnwidth]{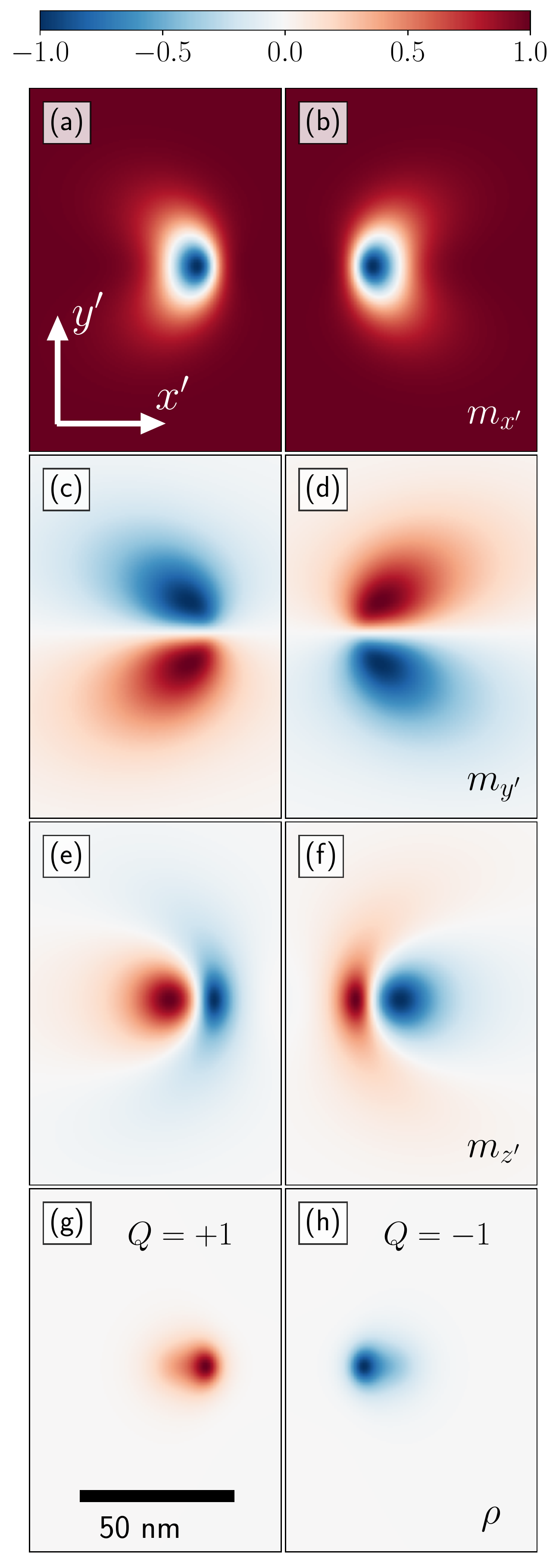}
	\caption{\label{Fig:ipskyr} In-plane skyrmions obtained using micromagnetic simulations with the
	in-plane anisotropy constant $K_u = 10^4\, {\rm J}/{\rm m}^3$. 
	The quantities are plotted in a rotated coordination system $(x', y', z')$ (see Fig.~\ref{Fig:dw90deg_dmi})
	so that the easy axis and outer magnetization are aligned with the $x'$-axis.
	The first three rows of the left and right column show the magnetization components 
	of an in-plane skyrmion with topological charged $Q=1$ and $Q=-1$, respectively; 
	(a) -- (b) $m_{x'}$, (c)--(d) $m_{y'}$, and (e)--(f) $m_{z'}$.
	In the fourth row is plotted the topological charge density.}
\end{figure}

Recent numerical analysis of thin magnetic layers with DMI and in-plane
uniaxial anisotropy with easy axis has revealed a possibility of 
the in-plane skyrmions existence~\cite{Moon:PRAppl_2019}. A stable in-plane skyrmion can be obtained
by rotating the out-of-plane magnetic texture with skyrmion into the easy-axis direction.
After relaxing the magnetic configuration one receives an in-plane skyrmion 
of a bean-like shape and out-of-plane core magnetization.
In contrast to the out-of-plane skyrmions, in-plane skyrmions of both skyrmion numbers $Q = \pm 1$
can coexist in one magnetizatic layer. 
Figure~\ref{Fig:ipskyr} shows in-plane skyrmions of both topological charges
obtained using micromagnetic simulations with DMI and $K_u = 10^4\, {\rm J}/{\rm m}^3$.
For higher resolution, we used discretization cells of size $\Delta x = \Delta y = \Delta z = 1\, {\rm nm}$.

Figures~\ref{Fig:ipskyr}(a)--(f) plot the magnetization components of the in-plane skyrmions.
In agreement with Fig.~\ref{Fig:dw90deg_dmi}, we plotted skyrmions in a rotated coordination 
system $(x', y', z')$ so that the easy axis and the outer magnetization 
are aligned with the horizontal axis, $x'$.
The left column depicts components of the in-plane skyrmion with topological charge $Q=1$,
while the right one plots magnetization for in-plane skyrmion with $Q=-1$.
In contrast to skyrmions in systems with perpendicular magnetic anisotropy, in-plane skyrmions
are not spherically symmetric. They show an axial symmetry with respect to the easy axis.
Moreover, their orientation is linked with their topological charge.
The in-plane skyrmions can be seen as a combination of an vortex and antivortex~\cite{Moon:PRAppl_2019},
known also as bimerons.
In case, of skyrmion with $Q=1$, the skyrmion core is formed by a vortex, while in the
opposite case, antivortex takes the role of the skyrmion core.
In Figures~\ref{Fig:ipskyr}(g)--(h) we map the topological charge density, defined as
\begin{equation}
	\rho(x,y) = \frac{1}{4\pi}\, {\bm m} \cdot \left( \pder{{\bm m}}{x} \times \pder{{\bm m}}{y} \right)
\label{Eq:top_density}
\end{equation}
The values of $\rho$ in Figs.~\ref{Fig:ipskyr} are normalized to the range of $\left\langle -1, 1 \right\rangle$. 
The maximum of the topological charge density is located between the vortex and antivortex of the skyrmion.
Unlike the out-of-plane skyrmions, the topological charge density of the in-plane skyrmion has a drop-like shape
elongated towards the skyrmion core.
The topological charge is then defined as $Q = \int {\rm d}x\, {\rm d}y\, \rho(x,y)$.

\begin{figure}
\centering
	\includegraphics[width=.8\columnwidth]{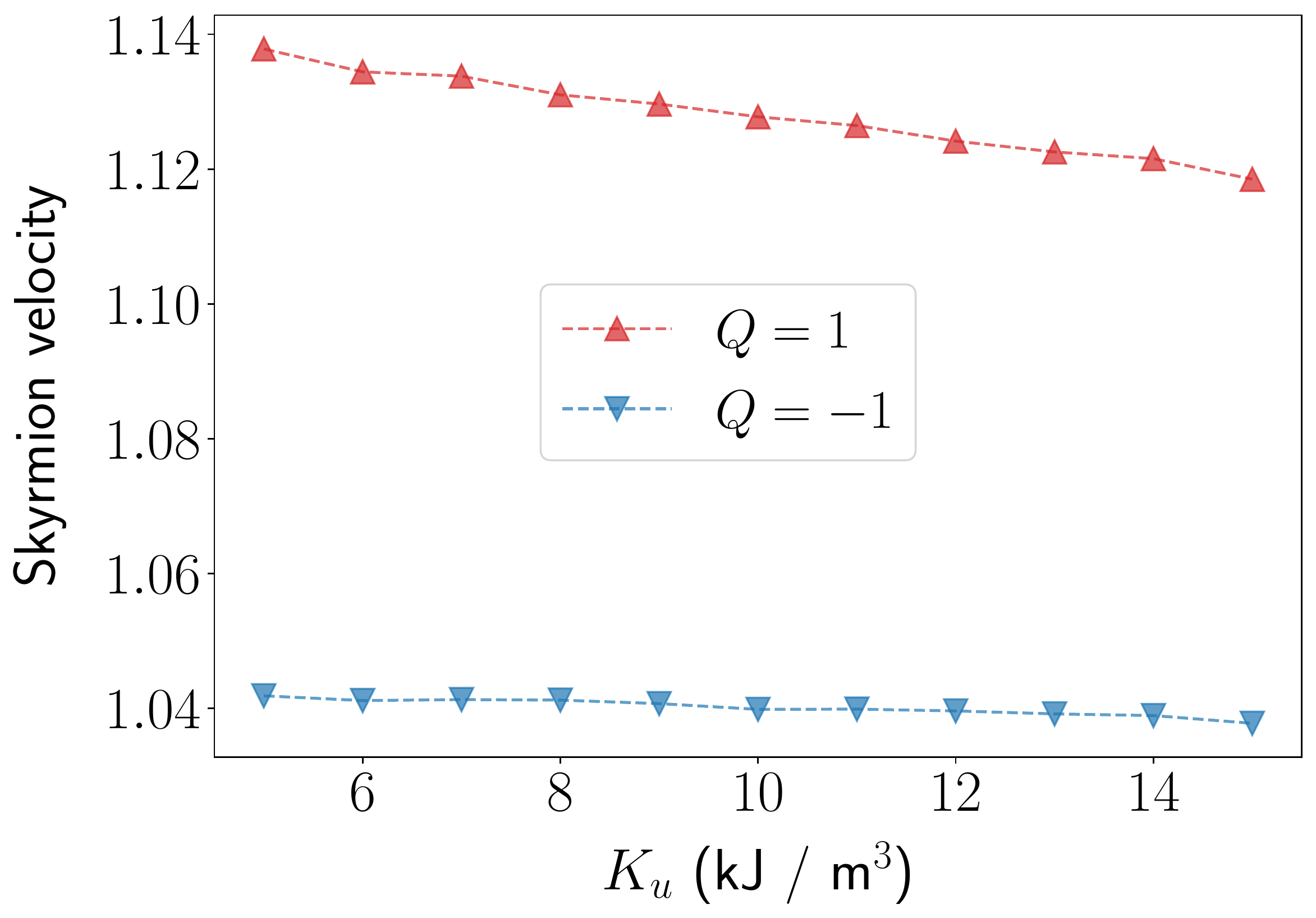}
	\caption{\label{Fig:sk_veloc_Ku} Skyrmion velocities in the central magnetic domain
	         of the sample shown in Fig.~\ref{Fig:dw90deg_dmi}(b) as a function of $K_u$ calculated for
	         both topological charges, $Q = \pm 1$, using micromagnetic simulations.
	         The current is applied along the $x$-axis, hence, under an angle $-45^\circ$ with respect to
	         the outer magnetization.
	         The skyrmion velocities are given in the units of $u$ defined by Eq.~\ref{Eq:u} with
	         parameters described below.}
\end{figure}
Let us now inspect the current-induced motion of the in-plane skyrmions in our system.
We assume an in-plane skyrmion nucleated in the middle of the sample shown in Fig.~\ref{Fig:dw90deg_dmi}(b).
To induce the skyrmion motion we assumed electric current flowing along the $x$-axis.
The resulting spin transfer torque has been modeled 
using the Li-Zhang torque~\cite{Li:PRL_2004,Li:PRB_2004} implemented in MuMax3~\cite{mumax3}.
First, we study the skyrmion velocity in the central magnetic domain of the sample
as a function of $K_u$ shown in Fig.~\ref{Fig:sk_veloc_Ku}.
We analyzed skyrmion velocity for both topological charges separately. 
The velocities in Fig.~\ref{Fig:sk_veloc_Ku} are plotted in the units of $u$ defined as
\begin{equation}
	u = \frac{\mu_{\rm B}\, P\, I}{2\, e\, M_{\rm s}\, (1 - \beta^2)}\,,
\label{Eq:u}
\end{equation}
where $\mu_{\rm B}$ is Bohr magneton, $P$ is the spin current polarization,
$I$ is the current density, and $e$ is the electron charge.
The values used in the calculations of the skyrmion velocity are
$P = 0.56$, $I = 10^{12}\, {\rm A}/{\rm m}^2$, and $\beta = 0.6$, which gives 
$|u| = 35.17\, {\rm m}/{\rm s}$.
Apparently, skyrmion velocities decrease with increasing $K_u$. 
More importantly, however, we notice a strong difference in the velocities of
skyrmions with opposite topological charge. In order to elucidate this difference,
we study the skyrmion velocities in more detail.
In the previous study by Moon {\em et al.}~\cite{Moon:PRAppl_2019}
it has been shown that the skyrmion velocity depends on current direction with respect to the easy axis.
Moreover, it has been shown that when the current flows along the easy axis,
skyrmions of both topological charges moves at the same velocities.
This is, however, not our case, since the current is applied under an angle of $-45^\circ$
with respect to the easy axis.
Thus we studied velocities of skyrmions of both topological charges as a function
of the current direction.

\begin{figure}
\centering
	\includegraphics[width=.8\columnwidth]{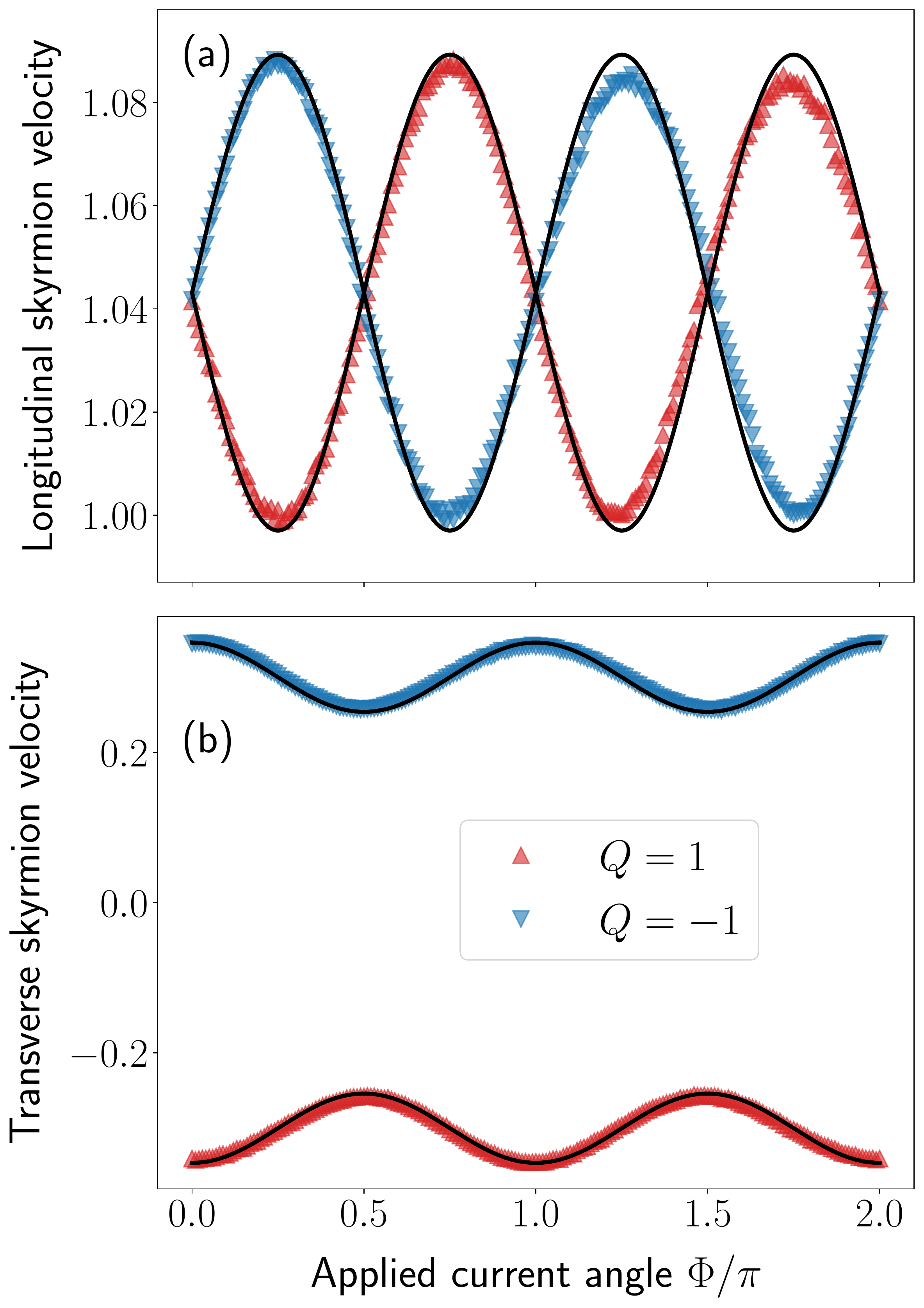}
	\caption{\label{Fig:sk_veloc_phi} Velocities of in-plane skyrmions 
	         in the central domain of sample shown in Fig.~\ref{Fig:dw90deg_dmi}(b)
	         as a function of the current direction.
	         The velocities are calculated for skyrmions of both topological charges, $Q = \pm 1$.
	         In Fig.~(a) we plot the longitudinal velocity component ($v_{\parallel}$), and panel (b)
	         shows the velocities perpendicular to the applied current direction ($v_{\perp}$).
	         The magnetic anisotropy parameter was $K_u = 10^{4}\, {\rm J}/{\rm m}^3$ 
	         and current density $I = 10^{12}\, {\rm A}/{\rm m}^2$, while the other
	         parameters are the same as on Fig.~\ref{Fig:sk_veloc_Ku}.
	         The current is applied under an angle $\Phi$ with respect to the $x'$-axis.
	         The velocities are plotted in the in the units of
	         $u$ given by Eq.~(\ref{Eq:u}).
		     The points show the skyrmion velocities estimated for the micromagnetic simulations, while
		     the black lines are corresponding skyrmion velocities calculated using the Thiele equation
		     (see Appendix~\ref{Sec:Thiele}).
		     Note, since the Thiele equation overestimates the skyrmion velocities, we shifted the calculated values
		     by a constant value, which is $-0.07$ in case of longitudinal velocities and $\pm 0.02$ in case of the
		     transverse velocities.}
\end{figure}
To estimate the skyrmion velocity with respect to the current direction 
the electric current of constant density $I = 10^{12}\, {\rm A}/{\rm m}^2$ 
is applied under an angle $\Phi$ with respect to the $x'$-axis (see Fig.~\ref{Fig:dw90deg_dmi}).
In order to separate the skyrmion Hall effect~\cite{Jiang:Nature_2017,Litzius:Nature_2017}, 
we split the velocity into two components;
the longitudinal component, $v_{\parallel}$, which is measured along the current direction, and
the transverse one, $v_\perp$, which is perpendicular to the current direction.
Triangles in Fig.~\ref{Fig:sk_veloc_phi} show the (a) longitudinal and (b) transverse skyrmion velocities 
calculated using micromagnetic simulations for skyrmions of both topological charges, $Q=\pm 1$.
We notice, that the longitudinal skyrmion velocity oscillate with angle $\Phi$ with period $\pi$.
The skyrmion velocity changes approximately in a range of $10\%$ of the average velocity.
Moreover, the velocities of skyrmions with different topological charges are shifted in phase by $\pi/2$. 
As a result, skyrmions of both topological charges move with the same 
longitudinal velocity when the electric current is oriented parallel or perpendicular to the easy axis.
The longitudinal velocities are equal in both current orientations. 
On the other hand, the perpendicular skyrmion velocity is slightly higher, when the electric current flows
parallel to the easy axis. 
In agreement with the theory of the skyrmion Hall effect, perpendicular velocities
differ in sign for opposite topological charges. 
Importantly, for any angle $\Phi$ different from $k\, \pi/2$ (for $k=0, 1, 2, \dots$), 
the skyrmion velocities of opposite $Q$ differ from each other.
The difference is maximum, when electric current is applied under an angle of $45^\circ$ with respect to the easy axis,
which is the case studied in Fig.~\ref{Fig:sk_veloc_Ku} as well as in the next section.

The skyrmion velocities described above are fully consistent with the Thiele equation~\cite{Thiele:PRL_1973}
studied also in Ref.~\cite{Moon:PRAppl_2019}. More details on the calculation of the skyrmion velocities
based on the Thiele equation can be found in Appendix~\ref{Sec:Thiele}.
The black lines in Fig.~\ref{Fig:sk_veloc_phi} are the results of the corresponding skyrmion velocities
using Eqs.~(\ref{Eqs:velocities}). Since the Thiele equation slightly overestimates the skyrmion velocities,
we shifted the calculated velocities by a value of $-0.07$ in Fig.~\ref{Fig:sk_veloc_phi}(a), and
by $\pm 0.02$ in Fig.~\ref{Fig:sk_veloc_phi}(b).
The overestimation of skyrmion velocities by the Thiele equation can be explained by additional dissipation mechanisms
related to dynamic variation of the skyrmion shape, which cannot be captured by the rigid shape approximation
used to derive the Thiele equation.
However, the skyrmion velocity oscillations and their amplitude are well captured by the Thiele equation. 

\section{Interaction of skyrmions with domain walls}
\label{Sec:SkyrmionsDW}

Up to now, we studied current-induced dynamics of in-plane skyrmions remaining in 
the central magnetic domain. Let us now inspect how the skyrmion motion is affected by 
the pinned $90^\circ$ magnetic domains.
Recently, interaction of skyrmions with DW has been studied in magnetic thin film with
perpendicular magnetic anisotropy~\cite{Song:APE_2020}. 
It has been shown, that a chiral DW can be considered as a guide for the skyrmion transport
preventing them from annihilation at the sample boundary.
Although we have not observed such an effect in the studied system with $90^\circ$ DWs, 
we show that $90^\circ$ DWs might have a significant effect on transport of the in-plane skyrmions.
It has been shown that the in-plane skyrmions can be
efficiently moved by means of the spin transfer torque or spin orbit torque~\cite{Moon:PRAppl_2019}.
In Sec.~\ref{SSec:current} we have demonstrated that applied electric current in the direction perpendicular to
the anisotropy boundaries has just a minor effect on the DW structure.
Thus we can use the spin transfer torque to study the common interaction between the $90^\circ$ DWs and 
the in-plane skyrmions.

\begin{figure}[htp]
	\centering
	\includegraphics[width=.99\columnwidth]{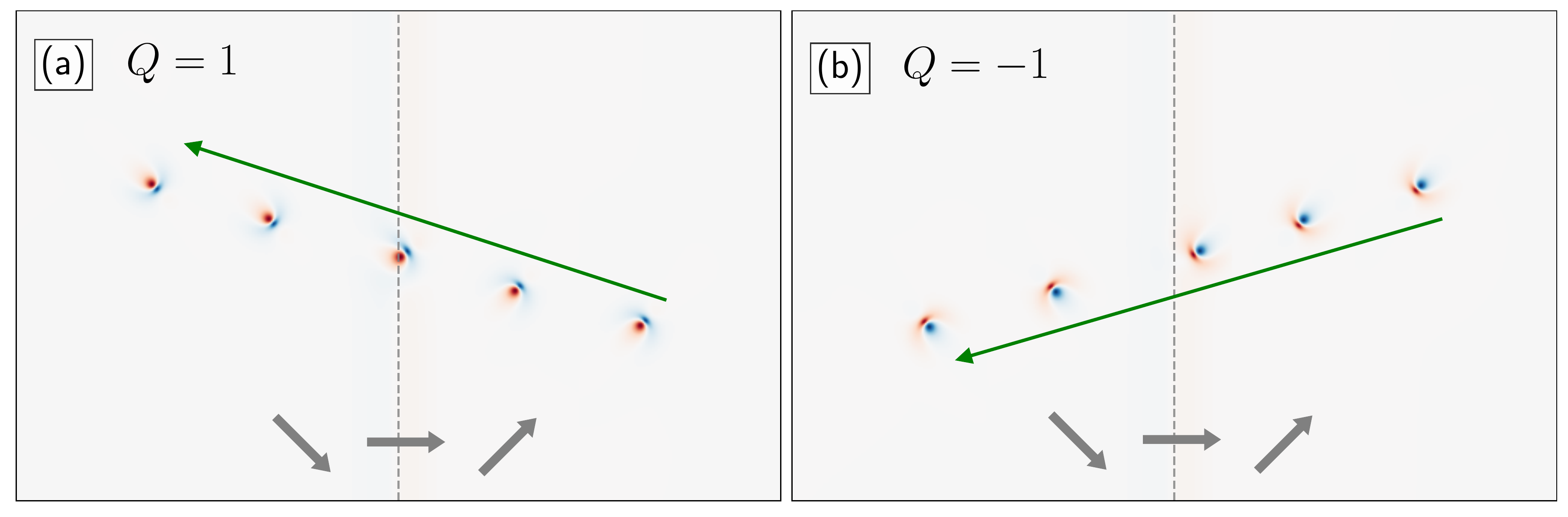}
	\caption{\label{Fig:traj_unch} In-plane skyrmions with topological charge 
	         (a) $Q=1$ and (b) $Q=-1$ passing an uncharged $90^\circ$ DW under the influence of applied current in the          
	         direction perpendicular to the domain wall. 
	         The colors show the magnetization $m_z$ components in the color scale identical 
	         to Fig.~\ref{Fig:ipskyr}.
	         The current density $I = 10^{12}\, {\rm A}/{\rm m}^2$ and anisotropy 
	         $K_u = 10^4\, {\rm J}/{\rm m}^3$ have been used in the simulations.
	         The single image shows skyrmions positions with time step approx.~$7\, {\rm ns}$, while
	         the green arrows give their direction. The vertical dashed lines marks the anisotropy boundary.
			 The grey thick arrows show the local magnetization directions.}
\end{figure}
First, we inspect how the in-plane skyrmion passes an uncharged $90^\circ$ magnetic DW.
As shown in the previous section, magnetization in the vicinity of an uncharged DW 
remains mainly in the layer's plane.
Considering the sample shown in Fig.~\ref{Fig:dw90deg_dmi}(a), we created an in-plane skyrmion in the center of the sample.
We assumed $D = 2\, {\rm mJ}/{\rm m}^2$ and $K_u = 10^4\, {\rm J}/{\rm m}^3$.
We move the skyrmions by applying a constant electric current of the density $I = 10^{12}\, {\rm A}/{\rm m}^2$ 
in the direction perpendicular to the DWs. For $I > 0$, the skyrmions move towards the left DW. 

In Figure~\ref{Fig:traj_unch}, we show motion of the in-plane skyrmions passing an uncharged magnetic DW.
The colors in Fig.~\ref{Fig:traj_unch} show the $z$-components of the reduced magnetization vector
in the color scale shown in Fig.~\ref{Fig:ipskyr}.
The anisotropy boundary is marked by the vertical dashed line and the local in-plane magnetization direction
is indicated by the thick grey arrows.
The single image shows skyrmion positions in different times with time step approx.~$7\, {\rm ns}$, while
the directions of moving skyrmions are given by the green arrows.
Fig.~\ref{Fig:traj_unch} covers $30\, {\rm ns}$ of the skyrmions dynamics.
Apparently, the skyrmion trajectories deviate from the horizontal direction due to the 
skyrmion Hall effect~\cite{Jiang:Nature_2017,Litzius:Nature_2017}.
Skyrmions with opposite topological charge are deflected in opposite directions.
As shown in Fig.~\ref{Fig:ipskyr}, the in-plane skyrmions are strictly oriented according to 
the local magnetization direction. Thus, when a skyrmion passes the domain wall it rotates to
the direction of magnetization of the neighboring domain. 
Importantly, the topological charge of the skyrmion remains unchanged, therefore,
also the direction of the skyrmion motion after passing the DW is the same.
For the opposite current direction with $I < 0$, the skyrmions 
move in the opposite direction towards the right DW (not shown).
The deflection of the skyrmion trajectories is also reversed.
Otherwise, the skyrmion transport through the right DW is analogical to the one 
described for the left DW.

\begin{figure}
	\centering
	\includegraphics[width=.99\columnwidth]{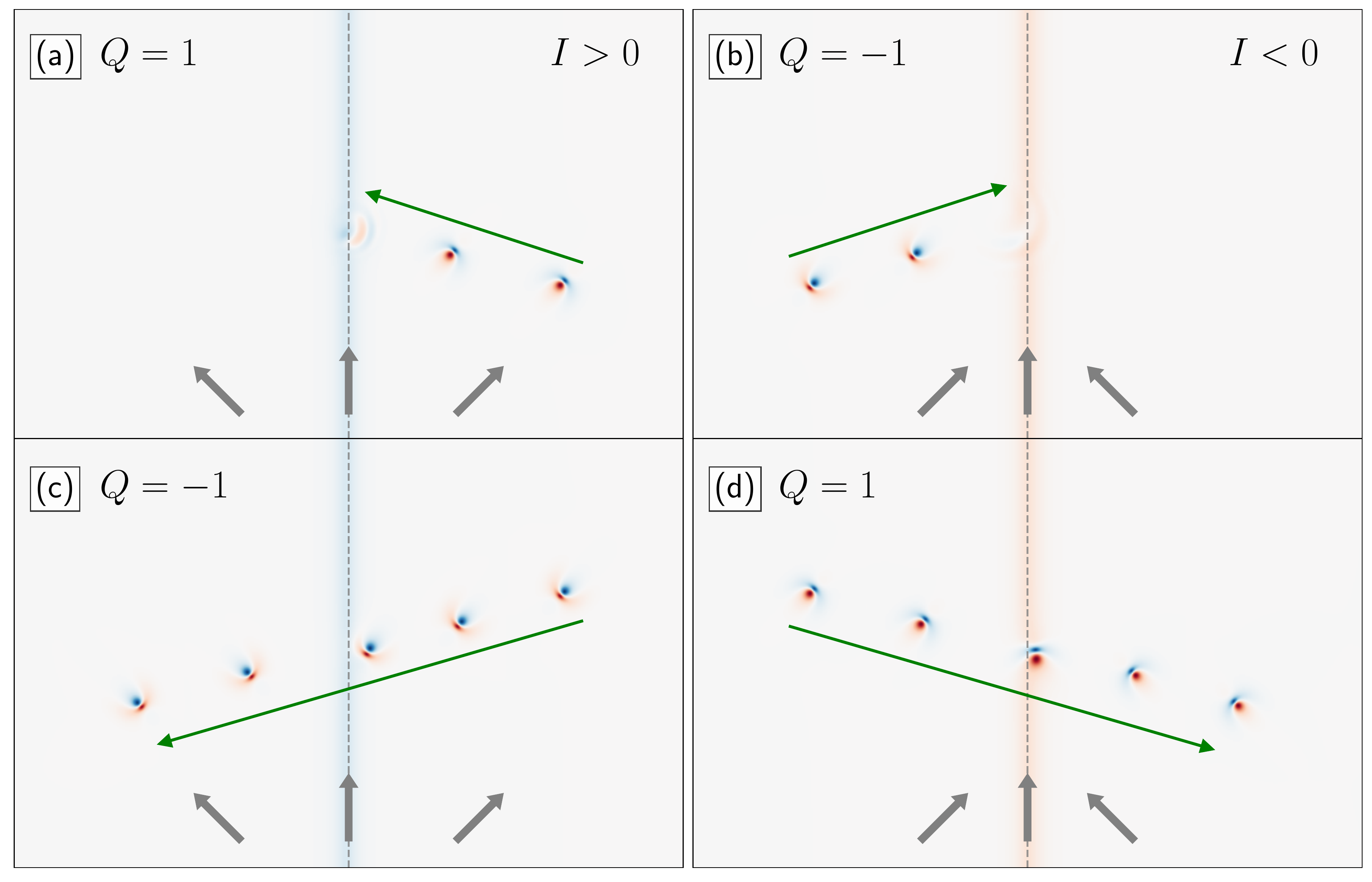}
	\caption{\label{Fig:traj_char} In-plane skyrmion with topological charged
	         (a), (d) $Q=1$ and (b), (c) $Q=-1$ passing a charged $90^\circ$ DW 
	         under the influence of applied current in the direction
	         perpendicular to the domain wall. 
	         The applied current density was (a), (c) $I = 10^{12}\, {\rm A}/{\rm m}^2$ and
	         (b), (d) $I = -10^{12}\, {\rm A}/{\rm m}^2$.
	         The other parameters are the same as in Fig.~\ref{Fig:traj_unch}.
	         The colors show the magnetization $m_z$ components in the color scale identical to Fig.~\ref{Fig:ipskyr}.
	         The single image shows skyrmions positions with time step approx.~$7\, {\rm ns}$, while
	         the green arrows give their direction.
	         The vertical dashed lines marks the anisotropy boundary.
			 The grey thick arrows show the local magnetization directions.}
\end{figure}
The situation is more complex in the case of charged DW. 
The summary of the skyrmion interaction with the $90^\circ$ charged magnetic DWs is shown in Fig.~\ref{Fig:traj_char}
analogically to Fig.~\ref{Fig:traj_unch}.
The simulation parameters are the same as in Fig.~\ref{Fig:traj_unch}.
In the left column of Fig.~\ref{Fig:traj_char}, we show how the in-plane skyrmions with different topological charges
interact with the left DW when $I > 0$.
While the skyrmion with $Q=-1$ passes the DW changing its orientation, the skyrmion with $Q=1$
is destroyed when hits the DW.
For $I < 0$, the skyrmions move towards the right DW, where the situation reverses. 
Namely, skyrmion with $Q=1$ transmits the DW, however, skyrmion with $Q=-1$ is annihilated.
We attribute this selective skyrmion annihilation at the localized $90^\circ$ DW to the
out-of-plane component induced by DMI. 
As follows out from our simulations, when the direction of the out-of-plane component of the DW
is in accord with the skyrmion core magnetization direction, the skyrmion passes through the DW.
In contrast, when the skyrmion core magnetization is opposite to the one of the DW out-of-plane magnetization 
component, the skyrmion becomes unstable. Inside an opposite DW, 
the skyrmion core shrinks down and the skyrmion annihilates.

Moreover, we find out that the effect of the charge selective skyrmion annihilation at the charged DWs
depends on the strength of the uniaxial anisotropy.
Namely, in our simulations we observe the effect when $K_u \gtrsim 7 \times 10^3\, {\rm J}/{\rm m}^3$
at the same value of $D = 2\, {\rm mJ}/{\rm m}^2$. 
Below this threshold skyrmions of both charges can transmit through both charged DWs.
In Fig.~\ref{Fig:Ku_dep} we show, that the magnitude of the out-of-plane magnetization component
in the DW center increases with increasing $K_u$. 
This suggests that the out-of-plane DW magnetization component is responsible for the skyrmion annihilation.
It also explains why such an effect is not observed in the case of the uncharged DWs, where the
out-of-plane magnetization is of one order of magnitude smaller.
In addition, we examined the stability of our results for a smaller Gilbert damping parameter.
Simulations with $\alpha = 0.03$ lead to the same qualitative results on skyrmion transmission and annihilation.

\section{In-plane skyrmion valve}
\label{Sec:Valve}

\begin{figure}[htp!]
	\centering
	\includegraphics[width=.99\columnwidth]{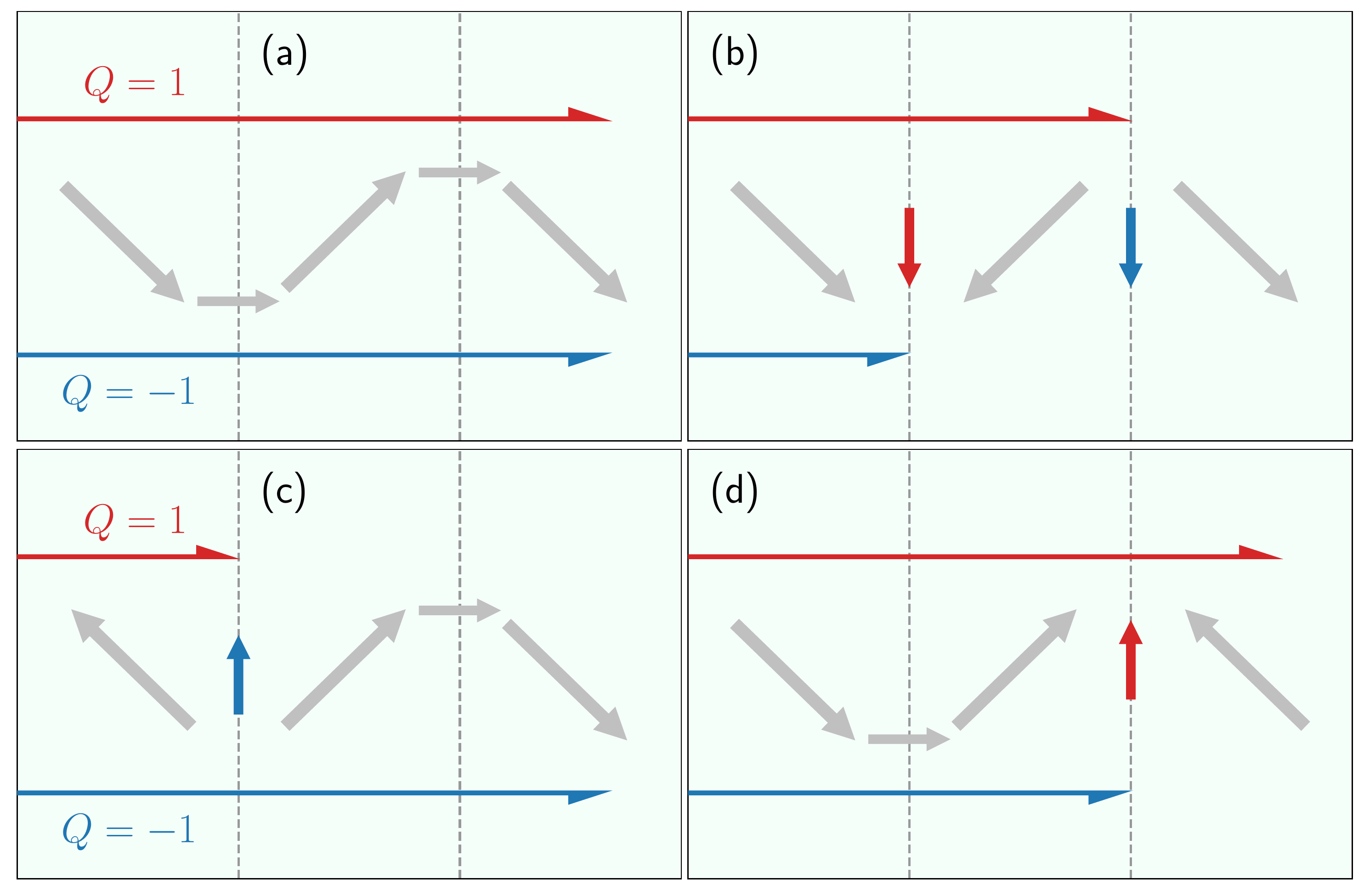}
	\caption{\label{Fig:device}
	Cartoon of an in-plane skyrmion valve based on two pinned $90^\circ$ magnetic DWs.
	Pictures shows three neighboring magnetic domain separated by $90^\circ$ magnetic DWs.
	The DWs are located at the gray dashed lines.
	The grey diagonal arrows show directions of magnetization in each of the domain.
	The smaller arrows show magnetizations in the centers of the DWs.
	The horizontal lines passing the structure depict the transmission of skyrmions
	with topological charge $Q = 1$ (red line) and $Q = -1$ (blue line).
	Figures (a) -- (d) present different magnetic configurations
	allowing or blocking transfer of in-plane skyrmions depending on their topological charge.}
\end{figure}
From the practical point of view, the effect of the topological charge dependent skyrmion annihilation can be utilized as
a skyrmion filter for construction of skyrmion-based logic gates.
This raises a question how to control the properties of the charged DWs and possibly
how to switch their effect on skyrmions on and off.
We have noticed that the out-of-plane magnetization of a charged DW must be large enough to be able to
annihilate a passing skyrmion. Thus changing the magnitude of the out-of-plane magnetization
might allow one to control the skyrmion flow through the DW.
As we have shown, the out-of-plane magnetization component can be varied in various ways.
On the one hand side, it can be manipulated by changing strength of the DMI or magnetic anisotropy.
On the other hand, it can be changed by an applied magnetic field, which can vary the magnetization angle
between two neighboring domains, and thus reduce the out-of-plane magnetization in the DW center.
These methods, however, do not affect only the $90^\circ$ magnetic DWs but also
the skyrmions, which can become unstable under distinct conditions~\cite{Moon:PRAppl_2019}.

Alternatively, one can consider the central magnetic domain in Fig.~\ref{Fig:dw90deg_dmi}
as a valve for the in-plane skyrmion flow. 
In Fig.~\ref{Fig:device} we present a concept of an in-plane skyrmion valve based on
two $90^\circ$ pinned magnetic DWs.
The cartoon in Fig.~\ref{Fig:device} shows a device consisted of three magnetic domains 
separated by two $90^\circ$ pinned DWs. The positions of the DWs are given by the
vertical dashed lines. The diagonal thick arrows correspond to directions of magnetization in each domain.
The smaller arrows located at the DW positions show the magnetization direction in the DW center.
Finally, the horizontal arrows passing the structure stand for the flow of skyrmions with
topological charge $Q=1$ (top arrow) and $Q=-1$ (bottom arrow).
Such a device has $2^3$ possible magnetic configurations.
In Fig.~\ref{Fig:device} we show its four basic functionalities.
Fig.~\ref{Fig:device}(a) shows a magnetic configuration featuring two uncharged magnetic DWs
corresponding to Fig.~\ref{Fig:dw90deg_dmi}(a).
In this configuration, skyrmions of both topological charges can 
freely move from one side of the sample to the other one.
However, when the magnetization in the central domain is switched,
both DWs turn to charged, as shown in Fig.~\ref{Fig:device}(b).
As a result, skyrmions of both topological charges will be annihilated at the two DWs
and none of them can pass the central domain.
Additionaly, if we want to introduce the topological charge selectivity into this scheme, we
need to combine charged and uncharged DWs. 
Two examples are shown in Figs.~\ref{Fig:device}(c) and (d).
In the first one, the magnetization of the left domain has been switched 
giving raise to a charged tail-to-tail domain wall.
This configuration allows just skyrmions with $Q=-1$ to pass the central domain. 
Oppositely, if the right domain magnetization is flipped, 
we obtain one charged head-to-head DW. 
Thus just skyrmions with $Q=1$ can be transferred to the other side of the sample.
Therefore, we suggest that such a device could fully control
the flow of the in-plane magnetic skyrmions.

To make this scheme operate in practice, an effective method of switching magnetization
selectively just in a specific magnetic stripe domain is necessary.
To this goal, one can employ the laser-induced magnetization precessional switching, 
which has been recently demonstrated in multiferroic composite BaTiO$_3$/CoFeB 
by Shelukhin {\em et al.}~\cite{Shelukhin:PRAppl_2020}
using femtosecond pump-probe technique with micrometer spatial resolution.
It has been shown that a femtosecond laser pulse can 
significantly reduce magnetoelastic coupling in a given domain. 
Consequently, with the assistance of 
ultrafast demagnetization~\cite{Battiato:PRL_2010,Carva:HMM_2017}
magnetization precessions can be triggered in the affected domain.
Importantly, when an external magnetic field is applied along the anisotropy hard axis
magnetization precessional switching~\cite{Schumacher:JAP_2003} can be achieved in the given domain  
without influencing its neighborhood.
In contrast to the previously described approaches,
the magnetic field is applied just for the period of magnetization switching.
Thus it does not affect the skyrmions passing the domain after the switching has been accomplished.
Moreover, the precessional magnetization reversal happens on a subnanosecond 
time scale~\cite{Shelukhin:PRAppl_2020}, 
which offers an ultrafast manipulation with the in-plane skyrmion valve.

\section{Conclusions}
\label{Sec:Discussion}

We have studied influence of interfacial DMI on the 
$90^\circ$ localized magnetic DWs formed in a thin magnetic layer with 
uniaxial in-plane magnetic anisotropy modified due to a ferroelectric substrate 
hosting stripe domains~\cite{Franke:PRL_2014}.
By means of micromagnetic simulations we have shown that, unlike in the uncharged magnetic DWs,
DMI induces significant out-of-plane magnetization component in the charged magnetic DWs.
The sign of the DMI-induced out-of-plane magnetization is opposite in case of the
head-to-head and tail-to-tail DW type. 
Consequently, we have studied magnetization dynamics of 
recently proposed in-plane magnetic skyrmions~\cite{Moon:PRAppl_2019}
in the system of magnetic stripe domains separated by $90^\circ$ magnetic DWs.
Particularly, we analyzed the longitudinal and transverse skyrmion velocities 
as a function of the applied current direction.
We demonstrate that the longitudinal skyrmion velocity also depends on the topological charge
which results in different velocities of skyrmions with opposite topological charges
when the electric current is applied under a general angle with respect to the easy axis.

Importantly, we have shown that the charged DWs have an important effect on the transport of
the in-plane skyrmions.
Namely, when an in-plane skyrmion passes a charged DW, it depends on its topological charge whether it
passes or becomes destroyed.
As follows from our simulations a tail-to-tail magnetic DW allows transmission of an in-plane skyrmion
of topological charge equal to $Q=-1$, however, becomes fatal to a skyrmion with $Q=1$.
On the other hand, a head-to-head DW is passable for a skyrmion with $Q=-1$ but annihilates skyrmion
with $Q=1$.
This process of skyrmion annihilation we attributed to the incompatibility of the out-of-plane magnetization
of the charged DW and the skyrmion core. 
When a skyrmion passes a pinned DW with large enough out-of-plane magnetization of opposite direction,
the skyrmion core becomes unstable and the skyrmion vanishes.

To achieve the coexistence of the pinned $90^\circ$ magnetic domain walls
and in-plane skyrmions one needs a composite featuring both
magnetic stripe domains and interfacial DMI. Although, to our best knowledge, such a device has
not been reported yet, a suitable candidate might be BaTiO$_3$/FM/Pt trilayer
with FM being a thin magnetic layer made of CoFe or CoFeB~\cite{Lahtinen:AdvMat_2011,Franke:PRL_2014}.
Once the material parameters fulfill the condition of in-plane skyrmion stability~\cite{Moon:PRAppl_2019},
they could be nucleated by known methods making use of 
spin injection~\cite{Sampaio:NatNanotech_2013,WOO-16}
or a laser pulse~\cite{Finazzi:PRL_2013,Gerlinger:APL_2021}.

Finally, we introduced a concept of a device based on two $90^\circ$ degrees pinned magnetic DWs 
allowing one to fully control the in-plane skyrmion flow. 
We suggest that making use of laser-induced spatially selective magnetization reversal~\cite{Shelukhin:PRAppl_2020}
such a device could be switched between different magnetic states
acting as a valve for in-plane magnetic skyrmions.

\section*{Acknowledgment}

This work was supported by the Czech Science Foundation (project no. 19-28594X).

\appendix

\section{Skyrmion velocities}
\label{Sec:Thiele}

Velocities of the in-plane skyrmions can be calculated using the Thiele equation~\cite{Thiele:PRL_1973,Iwasaki:NatComm_2013}.
Without any lost of generality, we assume here that the anisotropy easy axis is aligned with the $x$-axis.
In case of an in-plane skyrmion moved by the spin transfer torque~\cite{Li:PRL_2004,Li:PRB_2004} 
the Thiele equation leads to~\cite{Moon:PRAppl_2019}
\begin{subequations}
	\begin{align}
		-G\, \left(u_y - v_y \right) &= D_{xx}\, \left(\beta u_x - \alpha v_x \right)\,, \\
		G\, \left( u_x - v_x \right) &= D_{yy}\, \left(\beta u_y - \alpha v_y \right)\,,
	\end{align}
\label{Eqs:Thiele}
\end{subequations}
where $v_x$, and $v_y$ are the skyrmion velocities in the $x$, and $y$ direction, respectively.
Assuming the current density vector as ${\bm I} = I\, \left(\cos(\Phi), \sin(\Phi), 0 \right)$,
we obtain $u_x = u\, \cos(\Phi)$, and $u_y = u\, \sin(\Phi)$, where $u$ is given by Eq.~(\ref{Eq:u}).
Here, $\Phi$ is the angle of the applied current with respect to the $x$-axis.
Moreover, $G = -4\, \pi\, Q$, and 
$D_{\zeta\xi} = \int {\rm d}x\, {\rm d}y\, (\partial_\zeta {\bm m}) \cdot (\partial_\xi {\bm m})$
with ${\bm m} = {\bm m}(x, y)$ being the reduced magnetization vector.
From (\ref{Eqs:Thiele}) we obtain
\renewcommand*{\arraystretch}{1.5}
\begin{equation}
	\begin{pmatrix}
		v_x\\ v_y
	\end{pmatrix} = 
	\frac{u}{\Delta}\,
	{\cal D}
	\begin{pmatrix}
		\cos(\Phi)\\ \sin(\Phi)
	\end{pmatrix}
\end{equation}
where
\begin{equation}
	{\cal D} = 
	\begin{pmatrix}
		G^2 + \alpha\beta\, D_{xx} D_{yy} & G\, D_{yy}\, (\alpha - \beta)\\
		-G\, D_{xx}\, (\alpha - \beta) & G^2 + \alpha\beta\, D_{xx} D_{yy}
	\end{pmatrix}
\end{equation}
and	$\Delta = G^2 + \alpha\beta\, D_{xx} D_{yy}$.
Applying the rotation transformation, we obtain the longitudinal, $v_{\parallel}$, and
transverse, $v_{\perp}$, velocity components, which read
\begin{subequations}
	\begin{align}
	v_\parallel = \frac{u}{\Delta} \biggl\{ &G^2 + \alpha\beta\, D_{xx} D_{yy}\, - \notag \\
	&\frac{1}{2} \left( D_{xx} - D_{yy} \right)\, G\, (\alpha - \beta)\, \cos\left(2\Phi \right) \biggr\}\,, \\
	v_\perp =\, -\frac{1}{2}\, &\frac{u}{\Delta}\, G\, \left( \alpha - \beta \right) \times \notag \\
	&\left[ D_{xx} + D_{yy} + \left(D_{xx} - D_{yy} \right) \sin\left(2\Phi \right) \right]\,.
	\end{align}
\label{Eqs:velocities}
\end{subequations}

\bibliographystyle{apsrev4-1}

\begin{thebibliography}{49}%
\makeatletter
\providecommand \@ifxundefined [1]{%
 \@ifx{#1\undefined}
}%
\providecommand \@ifnum [1]{%
 \ifnum #1\expandafter \@firstoftwo
 \else \expandafter \@secondoftwo
 \fi
}%
\providecommand \@ifx [1]{%
 \ifx #1\expandafter \@firstoftwo
 \else \expandafter \@secondoftwo
 \fi
}%
\providecommand \natexlab [1]{#1}%
\providecommand \enquote  [1]{``#1''}%
\providecommand \bibnamefont  [1]{#1}%
\providecommand \bibfnamefont [1]{#1}%
\providecommand \citenamefont [1]{#1}%
\providecommand \href@noop [0]{\@secondoftwo}%
\providecommand \href [0]{\begingroup \@sanitize@url \@href}%
\providecommand \@href[1]{\@@startlink{#1}\@@href}%
\providecommand \@@href[1]{\endgroup#1\@@endlink}%
\providecommand \@sanitize@url [0]{\catcode `\\12\catcode `\$12\catcode
  `\&12\catcode `\#12\catcode `\^12\catcode `\_12\catcode `\%12\relax}%
\providecommand \@@startlink[1]{}%
\providecommand \@@endlink[0]{}%
\providecommand \url  [0]{\begingroup\@sanitize@url \@url }%
\providecommand \@url [1]{\endgroup\@href {#1}{\urlprefix }}%
\providecommand \urlprefix  [0]{URL }%
\providecommand \Eprint [0]{\href }%
\providecommand \doibase [0]{http://dx.doi.org/}%
\providecommand \selectlanguage [0]{\@gobble}%
\providecommand \bibinfo  [0]{\@secondoftwo}%
\providecommand \bibfield  [0]{\@secondoftwo}%
\providecommand \translation [1]{[#1]}%
\providecommand \BibitemOpen [0]{}%
\providecommand \bibitemStop [0]{}%
\providecommand \bibitemNoStop [0]{.\EOS\space}%
\providecommand \EOS [0]{\spacefactor3000\relax}%
\providecommand \BibitemShut  [1]{\csname bibitem#1\endcsname}%
\let\auto@bib@innerbib\@empty
\bibitem [{\citenamefont {Gradauskaite}\ \emph {et~al.}(2021)\citenamefont
  {Gradauskaite}, \citenamefont {Meisenheimer}, \citenamefont {Müller},
  \citenamefont {Heron},\ and\ \citenamefont
  {Trassin}}]{Gradauskaite:PSR_2021}%
  \BibitemOpen
  \bibfield  {author} {\bibinfo {author} {\bibfnamefont {E.}~\bibnamefont
  {Gradauskaite}}, \bibinfo {author} {\bibfnamefont {P.}~\bibnamefont
  {Meisenheimer}}, \bibinfo {author} {\bibfnamefont {M.}~\bibnamefont
  {Müller}}, \bibinfo {author} {\bibfnamefont {J.}~\bibnamefont {Heron}}, \
  and\ \bibinfo {author} {\bibfnamefont {M.}~\bibnamefont {Trassin}},\ }\href
  {\doibase doi:10.1515/psr-2019-0072} {\bibfield  {journal} {\bibinfo
  {journal} {Physical Sciences Reviews}\ }\textbf {\bibinfo {volume} {6}},\
  \bibinfo {pages} {20190072} (\bibinfo {year} {2021})}\BibitemShut {NoStop}%
\bibitem [{\citenamefont {Franke}\ \emph {et~al.}(2014)\citenamefont {Franke},
  \citenamefont {L\'opez~Gonz\'alez}, \citenamefont {H\"am\"al\"ainen},\ and\
  \citenamefont {van Dijken}}]{Franke:PRL_2014}%
  \BibitemOpen
  \bibfield  {author} {\bibinfo {author} {\bibfnamefont {K.~J.~A.}\
  \bibnamefont {Franke}}, \bibinfo {author} {\bibfnamefont {D.}~\bibnamefont
  {L\'opez~Gonz\'alez}}, \bibinfo {author} {\bibfnamefont {S.~J.}\ \bibnamefont
  {H\"am\"al\"ainen}}, \ and\ \bibinfo {author} {\bibfnamefont
  {S.}~\bibnamefont {van Dijken}},\ }\href {\doibase
  10.1103/PhysRevLett.112.017201} {\bibfield  {journal} {\bibinfo  {journal}
  {Phys. Rev. Lett.}\ }\textbf {\bibinfo {volume} {112}},\ \bibinfo {pages}
  {017201} (\bibinfo {year} {2014})}\BibitemShut {NoStop}%
\bibitem [{\citenamefont {Lebeugle}\ \emph {et~al.}(2009)\citenamefont
  {Lebeugle}, \citenamefont {Mougin}, \citenamefont {Viret}, \citenamefont
  {Colson},\ and\ \citenamefont {Ranno}}]{Lebeugle:PRL_2009}%
  \BibitemOpen
  \bibfield  {author} {\bibinfo {author} {\bibfnamefont {D.}~\bibnamefont
  {Lebeugle}}, \bibinfo {author} {\bibfnamefont {A.}~\bibnamefont {Mougin}},
  \bibinfo {author} {\bibfnamefont {M.}~\bibnamefont {Viret}}, \bibinfo
  {author} {\bibfnamefont {D.}~\bibnamefont {Colson}}, \ and\ \bibinfo {author}
  {\bibfnamefont {L.}~\bibnamefont {Ranno}},\ }\href {\doibase
  10.1103/PhysRevLett.103.257601} {\bibfield  {journal} {\bibinfo  {journal}
  {Phys. Rev. Lett.}\ }\textbf {\bibinfo {volume} {103}},\ \bibinfo {pages}
  {257601} (\bibinfo {year} {2009})}\BibitemShut {NoStop}%
\bibitem [{\citenamefont {Heron}\ \emph {et~al.}(2011)\citenamefont {Heron},
  \citenamefont {Trassin}, \citenamefont {Ashraf}, \citenamefont {Gajek},
  \citenamefont {He}, \citenamefont {Yang}, \citenamefont {Nikonov},
  \citenamefont {Chu}, \citenamefont {Salahuddin},\ and\ \citenamefont
  {Ramesh}}]{Heron:PRL_2011}%
  \BibitemOpen
  \bibfield  {author} {\bibinfo {author} {\bibfnamefont {J.~T.}\ \bibnamefont
  {Heron}}, \bibinfo {author} {\bibfnamefont {M.}~\bibnamefont {Trassin}},
  \bibinfo {author} {\bibfnamefont {K.}~\bibnamefont {Ashraf}}, \bibinfo
  {author} {\bibfnamefont {M.}~\bibnamefont {Gajek}}, \bibinfo {author}
  {\bibfnamefont {Q.}~\bibnamefont {He}}, \bibinfo {author} {\bibfnamefont
  {S.~Y.}\ \bibnamefont {Yang}}, \bibinfo {author} {\bibfnamefont {D.~E.}\
  \bibnamefont {Nikonov}}, \bibinfo {author} {\bibfnamefont {Y.-H.}\
  \bibnamefont {Chu}}, \bibinfo {author} {\bibfnamefont {S.}~\bibnamefont
  {Salahuddin}}, \ and\ \bibinfo {author} {\bibfnamefont {R.}~\bibnamefont
  {Ramesh}},\ }\href {\doibase 10.1103/PhysRevLett.107.217202} {\bibfield
  {journal} {\bibinfo  {journal} {Phys. Rev. Lett.}\ }\textbf {\bibinfo
  {volume} {107}},\ \bibinfo {pages} {217202} (\bibinfo {year}
  {2011})}\BibitemShut {NoStop}%
\bibitem [{\citenamefont {Lahtinen}\ \emph
  {et~al.}(2011{\natexlab{a}})\citenamefont {Lahtinen}, \citenamefont {Tuomi},\
  and\ \citenamefont {van Dijken}}]{Lahtinen:IEEETransMagn_2011}%
  \BibitemOpen
  \bibfield  {author} {\bibinfo {author} {\bibfnamefont {T.~H.~E.}\
  \bibnamefont {Lahtinen}}, \bibinfo {author} {\bibfnamefont {J.~O.}\
  \bibnamefont {Tuomi}}, \ and\ \bibinfo {author} {\bibfnamefont
  {S.}~\bibnamefont {van Dijken}},\ }\href {\doibase 10.1109/TMAG.2011.2143393}
  {\bibfield  {journal} {\bibinfo  {journal} {IEEE Trans. Magnetics}\ }\textbf
  {\bibinfo {volume} {47}},\ \bibinfo {pages} {3768} (\bibinfo {year}
  {2011}{\natexlab{a}})}\BibitemShut {NoStop}%
\bibitem [{\citenamefont {Lahtinen}\ \emph
  {et~al.}(2011{\natexlab{b}})\citenamefont {Lahtinen}, \citenamefont {Tuomi},\
  and\ \citenamefont {van Dijken}}]{Lahtinen:AdvMat_2011}%
  \BibitemOpen
  \bibfield  {author} {\bibinfo {author} {\bibfnamefont {T.~H.~E.}\
  \bibnamefont {Lahtinen}}, \bibinfo {author} {\bibfnamefont {J.~O.}\
  \bibnamefont {Tuomi}}, \ and\ \bibinfo {author} {\bibfnamefont
  {S.}~\bibnamefont {van Dijken}},\ }\href {\doibase 10.1002/adma.201100426}
  {\bibfield  {journal} {\bibinfo  {journal} {Adv. Mat.}\ }\textbf {\bibinfo
  {volume} {23}},\ \bibinfo {pages} {3187} (\bibinfo {year}
  {2011}{\natexlab{b}})}\BibitemShut {NoStop}%
\bibitem [{\citenamefont {Lahtinen}\ \emph {et~al.}(2012)\citenamefont
  {Lahtinen}, \citenamefont {Shirahata}, \citenamefont {Yao}, \citenamefont
  {Franke}, \citenamefont {Venkataiah}, \citenamefont {Taniyama},\ and\
  \citenamefont {van Dijken}}]{Lahtinen:APL_2012}%
  \BibitemOpen
  \bibfield  {author} {\bibinfo {author} {\bibfnamefont {T.~H.~E.}\
  \bibnamefont {Lahtinen}}, \bibinfo {author} {\bibfnamefont {Y.}~\bibnamefont
  {Shirahata}}, \bibinfo {author} {\bibfnamefont {L.}~\bibnamefont {Yao}},
  \bibinfo {author} {\bibfnamefont {K.~J.~A.}\ \bibnamefont {Franke}}, \bibinfo
  {author} {\bibfnamefont {G.}~\bibnamefont {Venkataiah}}, \bibinfo {author}
  {\bibfnamefont {T.}~\bibnamefont {Taniyama}}, \ and\ \bibinfo {author}
  {\bibfnamefont {S.}~\bibnamefont {van Dijken}},\ }\href {\doibase
  10.1063/1.4773482} {\bibfield  {journal} {\bibinfo  {journal} {Appl. Phys.
  Lett.}\ }\textbf {\bibinfo {volume} {101}},\ \bibinfo {pages} {262405}
  (\bibinfo {year} {2012})}\BibitemShut {NoStop}%
\bibitem [{\citenamefont {Chopdekar}\ \emph {et~al.}(2012)\citenamefont
  {Chopdekar}, \citenamefont {Malik}, \citenamefont {Fraile~Rodr\'{\i}guez},
  \citenamefont {Le~Guyader}, \citenamefont {Takamura}, \citenamefont {Scholl},
  \citenamefont {Stender}, \citenamefont {Schneider}, \citenamefont {Bernhard},
  \citenamefont {Nolting},\ and\ \citenamefont
  {Heyderman}}]{Chopdekar:PRB_2012}%
  \BibitemOpen
  \bibfield  {author} {\bibinfo {author} {\bibfnamefont {R.~V.}\ \bibnamefont
  {Chopdekar}}, \bibinfo {author} {\bibfnamefont {V.~K.}\ \bibnamefont
  {Malik}}, \bibinfo {author} {\bibfnamefont {A.}~\bibnamefont
  {Fraile~Rodr\'{\i}guez}}, \bibinfo {author} {\bibfnamefont {L.}~\bibnamefont
  {Le~Guyader}}, \bibinfo {author} {\bibfnamefont {Y.}~\bibnamefont
  {Takamura}}, \bibinfo {author} {\bibfnamefont {A.}~\bibnamefont {Scholl}},
  \bibinfo {author} {\bibfnamefont {D.}~\bibnamefont {Stender}}, \bibinfo
  {author} {\bibfnamefont {C.~W.}\ \bibnamefont {Schneider}}, \bibinfo {author}
  {\bibfnamefont {C.}~\bibnamefont {Bernhard}}, \bibinfo {author}
  {\bibfnamefont {F.}~\bibnamefont {Nolting}}, \ and\ \bibinfo {author}
  {\bibfnamefont {L.~J.}\ \bibnamefont {Heyderman}},\ }\href {\doibase
  10.1103/PhysRevB.86.014408} {\bibfield  {journal} {\bibinfo  {journal} {Phys.
  Rev. B}\ }\textbf {\bibinfo {volume} {86}},\ \bibinfo {pages} {014408}
  (\bibinfo {year} {2012})}\BibitemShut {NoStop}%
\bibitem [{\citenamefont {{Franke}}\ \emph {et~al.}(2015)\citenamefont
  {{Franke}}, \citenamefont {{Van de Wiele}}, \citenamefont {{Shirahata}},
  \citenamefont {{H{\"a}m{\"a}l{\"a}inen}}, \citenamefont {{Taniyama}},\ and\
  \citenamefont {{van Dijken}}}]{FRA-15}%
  \BibitemOpen
  \bibfield  {author} {\bibinfo {author} {\bibfnamefont {K.~J.~A.}\
  \bibnamefont {{Franke}}}, \bibinfo {author} {\bibfnamefont {B.}~\bibnamefont
  {{Van de Wiele}}}, \bibinfo {author} {\bibfnamefont {Y.}~\bibnamefont
  {{Shirahata}}}, \bibinfo {author} {\bibfnamefont {S.~J.}\ \bibnamefont
  {{H{\"a}m{\"a}l{\"a}inen}}}, \bibinfo {author} {\bibfnamefont
  {T.}~\bibnamefont {{Taniyama}}}, \ and\ \bibinfo {author} {\bibfnamefont
  {S.}~\bibnamefont {{van Dijken}}},\ }\href {\doibase
  10.1103/PhysRevX.5.011010} {\bibfield  {journal} {\bibinfo  {journal} {Phys.
  Rev. X}\ }\textbf {\bibinfo {volume} {5}},\ \bibinfo {eid} {011010} (\bibinfo
  {year} {2015})}\BibitemShut {NoStop}%
\bibitem [{\citenamefont {Van~de Wiele}\ \emph {et~al.}(2016)\citenamefont
  {Van~de Wiele}, \citenamefont {H\"am\"al\"ainen}, \citenamefont
  {Bal\'a\v{z}}, \citenamefont {Montoncello},\ and\ \citenamefont {van
  Dijken}}]{VanDeWiele:SciRep_2016}%
  \BibitemOpen
  \bibfield  {author} {\bibinfo {author} {\bibfnamefont {B.}~\bibnamefont
  {Van~de Wiele}}, \bibinfo {author} {\bibfnamefont {S.~J.}\ \bibnamefont
  {H\"am\"al\"ainen}}, \bibinfo {author} {\bibfnamefont {P.}~\bibnamefont
  {Bal\'a\v{z}}}, \bibinfo {author} {\bibfnamefont {F.}~\bibnamefont
  {Montoncello}}, \ and\ \bibinfo {author} {\bibfnamefont {S.}~\bibnamefont
  {van Dijken}},\ }\href {\doibase 10.1038/srep21330} {\bibfield  {journal}
  {\bibinfo  {journal} {Sci. Rep.}\ }\textbf {\bibinfo {volume} {6}},\ \bibinfo
  {pages} {21330} (\bibinfo {year} {2016})}\BibitemShut {NoStop}%
\bibitem [{\citenamefont {Bal\'a\v{z}}\ \emph {et~al.}(2018)\citenamefont
  {Bal\'a\v{z}}, \citenamefont {H\"am\"al\"ainen},\ and\ \citenamefont {van
  Dijken}}]{Balaz2018:PRB}%
  \BibitemOpen
  \bibfield  {author} {\bibinfo {author} {\bibfnamefont {P.}~\bibnamefont
  {Bal\'a\v{z}}}, \bibinfo {author} {\bibfnamefont {S.~J.}\ \bibnamefont
  {H\"am\"al\"ainen}}, \ and\ \bibinfo {author} {\bibfnamefont
  {S.}~\bibnamefont {van Dijken}},\ }\href {\doibase
  10.1103/PhysRevB.98.064417} {\bibfield  {journal} {\bibinfo  {journal} {Phys.
  Rev. B}\ }\textbf {\bibinfo {volume} {98}},\ \bibinfo {pages} {064417}
  (\bibinfo {year} {2018})}\BibitemShut {NoStop}%
\bibitem [{\citenamefont {{H{\"a}m{\"a}l{\"a}inen}}\ \emph
  {et~al.}(2017)\citenamefont {{H{\"a}m{\"a}l{\"a}inen}}, \citenamefont
  {{Brandl}}, \citenamefont {{Franke}}, \citenamefont {{Grundler}},\ and\
  \citenamefont {{van Dijken}}}]{HAM-17}%
  \BibitemOpen
  \bibfield  {author} {\bibinfo {author} {\bibfnamefont {S.~J.}\ \bibnamefont
  {{H{\"a}m{\"a}l{\"a}inen}}}, \bibinfo {author} {\bibfnamefont
  {F.}~\bibnamefont {{Brandl}}}, \bibinfo {author} {\bibfnamefont {K.~J.~A.}\
  \bibnamefont {{Franke}}}, \bibinfo {author} {\bibfnamefont {D.}~\bibnamefont
  {{Grundler}}}, \ and\ \bibinfo {author} {\bibfnamefont {S.}~\bibnamefont
  {{van Dijken}}},\ }\href {\doibase 10.1103/PhysRevApplied.8.014020}
  {\bibfield  {journal} {\bibinfo  {journal} {Phys. Rev. Applied}\ }\textbf
  {\bibinfo {volume} {8}},\ \bibinfo {eid} {014020} (\bibinfo {year}
  {2017})}\BibitemShut {NoStop}%
\bibitem [{\citenamefont {Shelukhin}\ \emph {et~al.}(2020)\citenamefont
  {Shelukhin}, \citenamefont {Pertsev}, \citenamefont {Scherbakov},
  \citenamefont {Kazenwadel}, \citenamefont {Kirilenko}, \citenamefont
  {H\"am\"al\"ainen}, \citenamefont {van Dijken},\ and\ \citenamefont
  {Kalashnikova}}]{Shelukhin:PRAppl_2020}%
  \BibitemOpen
  \bibfield  {author} {\bibinfo {author} {\bibfnamefont {L.~A.}~\bibnamefont
  {Shelukhin}}, \bibinfo {author} {\bibfnamefont {N.~A.}\ \bibnamefont
  {Pertsev}}, \bibinfo {author} {\bibfnamefont {A.~V.}~\bibnamefont {Scherbakov}},
  \bibinfo {author} {\bibfnamefont {D.~L.}~\bibnamefont {Kazenwadel}}, \bibinfo
  {author} {\bibfnamefont {D.~A.}~\bibnamefont {Kirilenko}}, \bibinfo {author}
  {\bibfnamefont {S.~J.}~\bibnamefont {H\"am\"al\"ainen}}, \bibinfo {author}
  {\bibfnamefont {S.}~\bibnamefont {van Dijken}}, \ and\ \bibinfo {author}
  {\bibfnamefont {A.~M.}~\bibnamefont {Kalashnikova}},\ }\href {\doibase
  10.1103/PhysRevApplied.14.034061} {\bibfield  {journal} {\bibinfo  {journal}
  {Phys. Rev. Applied}\ }\textbf {\bibinfo {volume} {14}},\ \bibinfo {pages}
  {034061} (\bibinfo {year} {2020})}\BibitemShut {NoStop}%
\bibitem [{\citenamefont {Dzyaloshinsky}(1958)}]{Dzyaloshinsky_1958}%
  \BibitemOpen
  \bibfield  {author} {\bibinfo {author} {\bibfnamefont {I.}~\bibnamefont
  {Dzyaloshinsky}},\ }\href {\doibase
  https://doi.org/10.1016/0022-3697(58)90076-3} {\bibfield  {journal} {\bibinfo
   {journal} {Journal of Physics and Chemistry of Solids}\ }\textbf {\bibinfo
  {volume} {4}},\ \bibinfo {pages} {241} (\bibinfo {year} {1958})}\BibitemShut
  {NoStop}%
\bibitem [{\citenamefont {Moriya}(1960)}]{Moriya_1960}%
  \BibitemOpen
  \bibfield  {author} {\bibinfo {author} {\bibfnamefont {T.}~\bibnamefont
  {Moriya}},\ }\href {\doibase 10.1103/PhysRev.120.91} {\bibfield  {journal}
  {\bibinfo  {journal} {Phys. Rev.}\ }\textbf {\bibinfo {volume} {120}},\
  \bibinfo {pages} {91} (\bibinfo {year} {1960})}\BibitemShut {NoStop}%
\bibitem [{\citenamefont {Crépieux}\ and\ \citenamefont
  {Lacroix}(1998)}]{Crepieux:JMMM_1998}%
  \BibitemOpen
  \bibfield  {author} {\bibinfo {author} {\bibfnamefont {A.}~\bibnamefont
  {Crépieux}}\ and\ \bibinfo {author} {\bibfnamefont {C.}~\bibnamefont
  {Lacroix}},\ }\href {\doibase https://doi.org/10.1016/S0304-8853(97)01044-5}
  {\bibfield  {journal} {\bibinfo  {journal} {J. Magn. Magn. Mater.}\ }\textbf
  {\bibinfo {volume} {182}},\ \bibinfo {pages} {341} (\bibinfo {year}
  {1998})}\BibitemShut {NoStop}%
\bibitem [{\citenamefont {Hrabec}\ \emph {et~al.}(2014)\citenamefont {Hrabec},
  \citenamefont {Porter}, \citenamefont {Wells}, \citenamefont {Benitez},
  \citenamefont {Burnell}, \citenamefont {McVitie}, \citenamefont {McGrouther},
  \citenamefont {Moore},\ and\ \citenamefont {Marrows}}]{Hrabec:PRB_2014}%
  \BibitemOpen
  \bibfield  {author} {\bibinfo {author} {\bibfnamefont {A.}~\bibnamefont
  {Hrabec}}, \bibinfo {author} {\bibfnamefont {N.~A.}\ \bibnamefont {Porter}},
  \bibinfo {author} {\bibfnamefont {A.}~\bibnamefont {Wells}}, \bibinfo
  {author} {\bibfnamefont {M.~J.}\ \bibnamefont {Benitez}}, \bibinfo {author}
  {\bibfnamefont {G.}~\bibnamefont {Burnell}}, \bibinfo {author} {\bibfnamefont
  {S.}~\bibnamefont {McVitie}}, \bibinfo {author} {\bibfnamefont
  {D.}~\bibnamefont {McGrouther}}, \bibinfo {author} {\bibfnamefont {T.~A.}\
  \bibnamefont {Moore}}, \ and\ \bibinfo {author} {\bibfnamefont {C.~H.}\
  \bibnamefont {Marrows}},\ }\href {\doibase 10.1103/PhysRevB.90.020402}
  {\bibfield  {journal} {\bibinfo  {journal} {Phys. Rev. B}\ }\textbf {\bibinfo
  {volume} {90}},\ \bibinfo {pages} {020402(R)} (\bibinfo {year}
  {2014})}\BibitemShut {NoStop}%
\bibitem [{\citenamefont {Je}\ \emph {et~al.}(2013)\citenamefont {Je},
  \citenamefont {Kim}, \citenamefont {Yoo}, \citenamefont {Min}, \citenamefont
  {Lee},\ and\ \citenamefont {Choe}}]{Je:PRB_2013}%
  \BibitemOpen
  \bibfield  {author} {\bibinfo {author} {\bibfnamefont {S.-G.}\ \bibnamefont
  {Je}}, \bibinfo {author} {\bibfnamefont {D.-H.}\ \bibnamefont {Kim}},
  \bibinfo {author} {\bibfnamefont {S.-C.}\ \bibnamefont {Yoo}}, \bibinfo
  {author} {\bibfnamefont {B.-C.}\ \bibnamefont {Min}}, \bibinfo {author}
  {\bibfnamefont {K.-J.}\ \bibnamefont {Lee}}, \ and\ \bibinfo {author}
  {\bibfnamefont {S.-B.}\ \bibnamefont {Choe}},\ }\href {\doibase
  10.1103/PhysRevB.88.214401} {\bibfield  {journal} {\bibinfo  {journal} {Phys.
  Rev. B}\ }\textbf {\bibinfo {volume} {88}},\ \bibinfo {pages} {214401}
  (\bibinfo {year} {2013})}\BibitemShut {NoStop}%
\bibitem [{\citenamefont {Cho}\ \emph {et~al.}(2015)\citenamefont {Cho},
  \citenamefont {Kim}, \citenamefont {Lee}, \citenamefont {Kim}, \citenamefont
  {Lavrijsen}, \citenamefont {Solignac}, \citenamefont {Yin}, \citenamefont
  {Han}, \citenamefont {van Hoof}, \citenamefont {Swagten}, \citenamefont
  {Koopmans},\ and\ \citenamefont {You}}]{Cho:NatComm_2015}%
  \BibitemOpen
  \bibfield  {author} {\bibinfo {author} {\bibfnamefont {J.}~\bibnamefont
  {Cho}}, \bibinfo {author} {\bibfnamefont {N.-H.}\ \bibnamefont {Kim}},
  \bibinfo {author} {\bibfnamefont {S.}~\bibnamefont {Lee}}, \bibinfo {author}
  {\bibfnamefont {J.-S.}\ \bibnamefont {Kim}}, \bibinfo {author} {\bibfnamefont
  {R.}~\bibnamefont {Lavrijsen}}, \bibinfo {author} {\bibfnamefont
  {A.}~\bibnamefont {Solignac}}, \bibinfo {author} {\bibfnamefont
  {Y.}~\bibnamefont {Yin}}, \bibinfo {author} {\bibfnamefont {D.-S.}\
  \bibnamefont {Han}}, \bibinfo {author} {\bibfnamefont {N.~J.~J.}\
  \bibnamefont {van Hoof}}, \bibinfo {author} {\bibfnamefont {H.~J.~M.}\
  \bibnamefont {Swagten}}, \bibinfo {author} {\bibfnamefont {B.}~\bibnamefont
  {Koopmans}}, \ and\ \bibinfo {author} {\bibfnamefont {C.-Y.}\ \bibnamefont
  {You}},\ }\href {\doibase 10.1038/ncomms8635} {\bibfield  {journal} {\bibinfo
   {journal} {Nature Commun.}\ }\textbf {\bibinfo {volume} {6}},\ \bibinfo
  {pages} {7635} (\bibinfo {year} {2015})}\BibitemShut {NoStop}%
\bibitem [{\citenamefont {Kim}\ \emph {et~al.}(2018)\citenamefont {Kim},
  \citenamefont {Moon}, \citenamefont {Kerber}, \citenamefont {Nothhelfer},\
  and\ \citenamefont {Everschor-Sitte}}]{Kim:PRB_2018}%
  \BibitemOpen
  \bibfield  {author} {\bibinfo {author} {\bibfnamefont {K.-W.}\ \bibnamefont
  {Kim}}, \bibinfo {author} {\bibfnamefont {K.-W.}\ \bibnamefont {Moon}},
  \bibinfo {author} {\bibfnamefont {N.}~\bibnamefont {Kerber}}, \bibinfo
  {author} {\bibfnamefont {J.}~\bibnamefont {Nothhelfer}}, \ and\ \bibinfo
  {author} {\bibfnamefont {K.}~\bibnamefont {Everschor-Sitte}},\ }\href
  {\doibase 10.1103/PhysRevB.97.224427} {\bibfield  {journal} {\bibinfo
  {journal} {Phys. Rev. B}\ }\textbf {\bibinfo {volume} {97}},\ \bibinfo
  {pages} {224427} (\bibinfo {year} {2018})}\BibitemShut {NoStop}%
\bibitem [{\citenamefont {Thiaville}\ \emph {et~al.}(2012)\citenamefont
  {Thiaville}, \citenamefont {Rohart}, \citenamefont {Ju{\'{e}}}, \citenamefont
  {Cros},\ and\ \citenamefont {Fert}}]{Thiaville:EPL_2012}%
  \BibitemOpen
  \bibfield  {author} {\bibinfo {author} {\bibfnamefont {A.}~\bibnamefont
  {Thiaville}}, \bibinfo {author} {\bibfnamefont {S.}~\bibnamefont {Rohart}},
  \bibinfo {author} {\bibfnamefont {{\'{E}}.}~\bibnamefont {Ju{\'{e}}}},
  \bibinfo {author} {\bibfnamefont {V.}~\bibnamefont {Cros}}, \ and\ \bibinfo
  {author} {\bibfnamefont {A.}~\bibnamefont {Fert}},\ }\href {\doibase
  10.1209/0295-5075/100/57002} {\bibfield  {journal} {\bibinfo  {journal}
  {{EPL} (Europhysics Letters)}\ }\textbf {\bibinfo {volume} {100}},\ \bibinfo
  {pages} {57002} (\bibinfo {year} {2012})}\BibitemShut {NoStop}%
\bibitem [{\citenamefont {Garcia}\ \emph {et~al.}(2021)\citenamefont {Garcia},
  \citenamefont {Fassatoui}, \citenamefont {Bonfim}, \citenamefont {Vogel},
  \citenamefont {Thiaville},\ and\ \citenamefont {Pizzini}}]{Garcia:PRB_2021}%
  \BibitemOpen
  \bibfield  {author} {\bibinfo {author} {\bibfnamefont {J.~P.~n.}\
  \bibnamefont {Garcia}}, \bibinfo {author} {\bibfnamefont {A.}~\bibnamefont
  {Fassatoui}}, \bibinfo {author} {\bibfnamefont {M.}~\bibnamefont {Bonfim}},
  \bibinfo {author} {\bibfnamefont {J.}~\bibnamefont {Vogel}}, \bibinfo
  {author} {\bibfnamefont {A.}~\bibnamefont {Thiaville}}, \ and\ \bibinfo
  {author} {\bibfnamefont {S.}~\bibnamefont {Pizzini}},\ }\href {\doibase
  10.1103/PhysRevB.104.014405} {\bibfield  {journal} {\bibinfo  {journal}
  {Phys. Rev. B}\ }\textbf {\bibinfo {volume} {104}},\ \bibinfo {pages}
  {014405} (\bibinfo {year} {2021})}\BibitemShut {NoStop}%
\bibitem [{\citenamefont {Li}\ \emph {et~al.}(2015)\citenamefont {Li},
  \citenamefont {Liu}, \citenamefont {Li},\ and\ \citenamefont
  {He}}]{Li:JAP_2015}%
  \BibitemOpen
  \bibfield  {author} {\bibinfo {author} {\bibfnamefont {Z.-D.}\ \bibnamefont
  {Li}}, \bibinfo {author} {\bibfnamefont {F.}~\bibnamefont {Liu}}, \bibinfo
  {author} {\bibfnamefont {Q.-Y.}\ \bibnamefont {Li}}, \ and\ \bibinfo {author}
  {\bibfnamefont {P.~B.}\ \bibnamefont {He}},\ }\href {\doibase
  10.1063/1.4919676} {\bibfield  {journal} {\bibinfo  {journal} {Journal of
  Applied Physics}\ }\textbf {\bibinfo {volume} {117}},\ \bibinfo {pages}
  {173906} (\bibinfo {year} {2015})},\ \Eprint
  {http://arxiv.org/abs/https://doi.org/10.1063/1.4919676}
  {https://doi.org/10.1063/1.4919676} \BibitemShut {NoStop}%
\bibitem [{\citenamefont {Moon}\ \emph {et~al.}(2019)\citenamefont {Moon},
  \citenamefont {Yoon}, \citenamefont {Kim},\ and\ \citenamefont
  {Hwang}}]{Moon:PRAppl_2019}%
  \BibitemOpen
  \bibfield  {author} {\bibinfo {author} {\bibfnamefont {K.-W.}\ \bibnamefont
  {Moon}}, \bibinfo {author} {\bibfnamefont {J.}~\bibnamefont {Yoon}}, \bibinfo
  {author} {\bibfnamefont {C.}~\bibnamefont {Kim}}, \ and\ \bibinfo {author}
  {\bibfnamefont {C.}~\bibnamefont {Hwang}},\ }\href {\doibase
  10.1103/PhysRevApplied.12.064054} {\bibfield  {journal} {\bibinfo  {journal}
  {Phys. Rev. Applied}\ }\textbf {\bibinfo {volume} {12}},\ \bibinfo {pages}
  {064054} (\bibinfo {year} {2019})}\BibitemShut {NoStop}%
\bibitem [{\citenamefont {Zhang}\ \emph {et~al.}(2015)\citenamefont {Zhang},
  \citenamefont {Ezawa},\ and\ \citenamefont {Zhou}}]{Zhang:SciRep_2015}%
  \BibitemOpen
  \bibfield  {author} {\bibinfo {author} {\bibfnamefont {X.}~\bibnamefont
  {Zhang}}, \bibinfo {author} {\bibfnamefont {M.}~\bibnamefont {Ezawa}}, \ and\
  \bibinfo {author} {\bibfnamefont {Y.}~\bibnamefont {Zhou}},\ }\href {\doibase
  10.1038/srep09400} {\bibfield  {journal} {\bibinfo  {journal} {Scientific
  Reports}\ }\textbf {\bibinfo {volume} {5}},\ \bibinfo {pages} {9400}
  (\bibinfo {year} {2015})}\BibitemShut {NoStop}%
\bibitem [{\citenamefont {Kharkov}\ \emph {et~al.}(2017)\citenamefont
  {Kharkov}, \citenamefont {Sushkov},\ and\ \citenamefont
  {Mostovoy}}]{Kharkov:PRL_2017}%
  \BibitemOpen
  \bibfield  {author} {\bibinfo {author} {\bibfnamefont {Y.~A.}\ \bibnamefont
  {Kharkov}}, \bibinfo {author} {\bibfnamefont {O.~P.}\ \bibnamefont
  {Sushkov}}, \ and\ \bibinfo {author} {\bibfnamefont {M.}~\bibnamefont
  {Mostovoy}},\ }\href {\doibase 10.1103/PhysRevLett.119.207201} {\bibfield
  {journal} {\bibinfo  {journal} {Phys. Rev. Lett.}\ }\textbf {\bibinfo
  {volume} {119}},\ \bibinfo {pages} {207201} (\bibinfo {year}
  {2017})}\BibitemShut {NoStop}%
\bibitem [{\citenamefont {G\"obel}\ \emph {et~al.}(2019)\citenamefont
  {G\"obel}, \citenamefont {Mook}, \citenamefont {Henk}, \citenamefont
  {Mertig},\ and\ \citenamefont {Tretiakov}}]{Gobel:PRB_2019}%
  \BibitemOpen
  \bibfield  {author} {\bibinfo {author} {\bibfnamefont {B.}~\bibnamefont
  {G\"obel}}, \bibinfo {author} {\bibfnamefont {A.}~\bibnamefont {Mook}},
  \bibinfo {author} {\bibfnamefont {J.}~\bibnamefont {Henk}}, \bibinfo {author}
  {\bibfnamefont {I.}~\bibnamefont {Mertig}}, \ and\ \bibinfo {author}
  {\bibfnamefont {O.~A.}\ \bibnamefont {Tretiakov}},\ }\href {\doibase
  10.1103/PhysRevB.99.060407} {\bibfield  {journal} {\bibinfo  {journal} {Phys.
  Rev. B}\ }\textbf {\bibinfo {volume} {99}},\ \bibinfo {pages} {060407(R)}
  (\bibinfo {year} {2019})}\BibitemShut {NoStop}%
\bibitem [{\citenamefont {Li}\ \emph {et~al.}(2020)\citenamefont {Li},
  \citenamefont {Shen}, \citenamefont {Bai}, \citenamefont {Wang},
  \citenamefont {Zhang}, \citenamefont {Xia}, \citenamefont {Ezawa},
  \citenamefont {Tretiakov}, \citenamefont {Xu}, \citenamefont {Mruczkiewicz},
  \citenamefont {Krawczyk}, \citenamefont {Xu}, \citenamefont {Evans},
  \citenamefont {Chantrell},\ and\ \citenamefont {Zhou}}]{Li:NPJCM_2020}%
  \BibitemOpen
  \bibfield  {author} {\bibinfo {author} {\bibfnamefont {X.}~\bibnamefont
  {Li}}, \bibinfo {author} {\bibfnamefont {L.}~\bibnamefont {Shen}}, \bibinfo
  {author} {\bibfnamefont {Y.}~\bibnamefont {Bai}}, \bibinfo {author}
  {\bibfnamefont {J.}~\bibnamefont {Wang}}, \bibinfo {author} {\bibfnamefont
  {X.}~\bibnamefont {Zhang}}, \bibinfo {author} {\bibfnamefont
  {J.}~\bibnamefont {Xia}}, \bibinfo {author} {\bibfnamefont {M.}~\bibnamefont
  {Ezawa}}, \bibinfo {author} {\bibfnamefont {O.~A.}\ \bibnamefont
  {Tretiakov}}, \bibinfo {author} {\bibfnamefont {X.}~\bibnamefont {Xu}},
  \bibinfo {author} {\bibfnamefont {M.}~\bibnamefont {Mruczkiewicz}}, \bibinfo
  {author} {\bibfnamefont {M.}~\bibnamefont {Krawczyk}}, \bibinfo {author}
  {\bibfnamefont {Y.}~\bibnamefont {Xu}}, \bibinfo {author} {\bibfnamefont
  {R.~F.~L.}\ \bibnamefont {Evans}}, \bibinfo {author} {\bibfnamefont {R.~W.}\
  \bibnamefont {Chantrell}}, \ and\ \bibinfo {author} {\bibfnamefont
  {Y.}~\bibnamefont {Zhou}},\ }\href {\doibase 10.1038/s41524-020-00435-y}
  {\bibfield  {journal} {\bibinfo  {journal} {npj Computational Materials}\
  }\textbf {\bibinfo {volume} {6}},\ \bibinfo {pages} {169} (\bibinfo {year}
  {2020})}\BibitemShut {NoStop}%
\bibitem [{\citenamefont {Jiang}\ \emph {et~al.}(2017)\citenamefont {Jiang},
  \citenamefont {Zhang}, \citenamefont {Yu}, \citenamefont {Zhang},
  \citenamefont {Wang}, \citenamefont {Benjamin~Jungfleisch}, \citenamefont
  {Pearson}, \citenamefont {Cheng}, \citenamefont {Heinonen}, \citenamefont
  {Wang}, \citenamefont {Zhou}, \citenamefont {Hoffmann},\ and\ \citenamefont
  {te~Velthuis}}]{Jiang:Nature_2017}%
  \BibitemOpen
  \bibfield  {author} {\bibinfo {author} {\bibfnamefont {W.}~\bibnamefont
  {Jiang}}, \bibinfo {author} {\bibfnamefont {X.}~\bibnamefont {Zhang}},
  \bibinfo {author} {\bibfnamefont {G.}~\bibnamefont {Yu}}, \bibinfo {author}
  {\bibfnamefont {W.}~\bibnamefont {Zhang}}, \bibinfo {author} {\bibfnamefont
  {X.}~\bibnamefont {Wang}}, \bibinfo {author} {\bibfnamefont {M.}~\bibnamefont
  {Benjamin~Jungfleisch}}, \bibinfo {author} {\bibfnamefont {J.~E.}\
  \bibnamefont {Pearson}}, \bibinfo {author} {\bibfnamefont {X.}~\bibnamefont
  {Cheng}}, \bibinfo {author} {\bibfnamefont {O.}~\bibnamefont {Heinonen}},
  \bibinfo {author} {\bibfnamefont {K.~L.}\ \bibnamefont {Wang}}, \bibinfo
  {author} {\bibfnamefont {Y.}~\bibnamefont {Zhou}}, \bibinfo {author}
  {\bibfnamefont {A.}~\bibnamefont {Hoffmann}}, \ and\ \bibinfo {author}
  {\bibfnamefont {S.~G.~E.}\ \bibnamefont {te~Velthuis}},\ }\href {\doibase
  10.1038/nphys3883} {\bibfield  {journal} {\bibinfo  {journal} {Nature
  Physics}\ }\textbf {\bibinfo {volume} {13}},\ \bibinfo {pages} {162}
  (\bibinfo {year} {2017})}\BibitemShut {NoStop}%
\bibitem [{\citenamefont {Litzius}\ \emph {et~al.}(2017)\citenamefont
  {Litzius}, \citenamefont {Lemesh}, \citenamefont {Kr{\"u}ger}, \citenamefont
  {Bassirian}, \citenamefont {Caretta}, \citenamefont {Richter}, \citenamefont
  {B{\"u}ttner}, \citenamefont {Sato}, \citenamefont {Tretiakov}, \citenamefont
  {F{\"o}rster}, \citenamefont {Reeve}, \citenamefont {Weigand}, \citenamefont
  {Bykova}, \citenamefont {Stoll}, \citenamefont {Sch{\"u}tz}, \citenamefont
  {Beach},\ and\ \citenamefont {Kl{\"a}ui}}]{Litzius:Nature_2017}%
  \BibitemOpen
  \bibfield  {author} {\bibinfo {author} {\bibfnamefont {K.}~\bibnamefont
  {Litzius}}, \bibinfo {author} {\bibfnamefont {I.}~\bibnamefont {Lemesh}},
  \bibinfo {author} {\bibfnamefont {B.}~\bibnamefont {Kr{\"u}ger}}, \bibinfo
  {author} {\bibfnamefont {P.}~\bibnamefont {Bassirian}}, \bibinfo {author}
  {\bibfnamefont {L.}~\bibnamefont {Caretta}}, \bibinfo {author} {\bibfnamefont
  {K.}~\bibnamefont {Richter}}, \bibinfo {author} {\bibfnamefont
  {F.}~\bibnamefont {B{\"u}ttner}}, \bibinfo {author} {\bibfnamefont
  {K.}~\bibnamefont {Sato}}, \bibinfo {author} {\bibfnamefont {O.~A.}\
  \bibnamefont {Tretiakov}}, \bibinfo {author} {\bibfnamefont {J.}~\bibnamefont
  {F{\"o}rster}}, \bibinfo {author} {\bibfnamefont {R.~M.}\ \bibnamefont
  {Reeve}}, \bibinfo {author} {\bibfnamefont {M.}~\bibnamefont {Weigand}},
  \bibinfo {author} {\bibfnamefont {I.}~\bibnamefont {Bykova}}, \bibinfo
  {author} {\bibfnamefont {H.}~\bibnamefont {Stoll}}, \bibinfo {author}
  {\bibfnamefont {G.}~\bibnamefont {Sch{\"u}tz}}, \bibinfo {author}
  {\bibfnamefont {G.~S.~D.}\ \bibnamefont {Beach}}, \ and\ \bibinfo {author}
  {\bibfnamefont {M.}~\bibnamefont {Kl{\"a}ui}},\ }\href {\doibase
  10.1038/nphys4000} {\bibfield  {journal} {\bibinfo  {journal} {Nature
  Physics}\ }\textbf {\bibinfo {volume} {13}},\ \bibinfo {pages} {170}
  (\bibinfo {year} {2017})}\BibitemShut {NoStop}%
\bibitem [{\citenamefont {Vansteenkiste}\ \emph {et~al.}(2014)\citenamefont
  {Vansteenkiste}, \citenamefont {Leliaert}, \citenamefont {Dvornik},
  \citenamefont {Helsen}, \citenamefont {Garcia-Sanchez},\ and\ \citenamefont
  {Waeyenberge}}]{mumax3}%
  \BibitemOpen
  \bibfield  {author} {\bibinfo {author} {\bibfnamefont {A.}~\bibnamefont
  {Vansteenkiste}}, \bibinfo {author} {\bibfnamefont {J.}~\bibnamefont
  {Leliaert}}, \bibinfo {author} {\bibfnamefont {M.}~\bibnamefont {Dvornik}},
  \bibinfo {author} {\bibfnamefont {M.}~\bibnamefont {Helsen}}, \bibinfo
  {author} {\bibfnamefont {F.}~\bibnamefont {Garcia-Sanchez}}, \ and\ \bibinfo
  {author} {\bibfnamefont {B.~V.}\ \bibnamefont {Waeyenberge}},\ }\href
  {\doibase 10.1063/1.4899186} {\bibfield  {journal} {\bibinfo  {journal} {AIP
  Advances}\ }\textbf {\bibinfo {volume} {4}},\ \bibinfo {pages} {107133}
  (\bibinfo {year} {2014})}\BibitemShut {NoStop}%
\bibitem [{\citenamefont {Piao}\ \emph {et~al.}(2011)\citenamefont {Piao},
  \citenamefont {Choi}, \citenamefont {Shim}, \citenamefont {Kim},\ and\
  \citenamefont {You}}]{Piao:APL_2011}%
  \BibitemOpen
  \bibfield  {author} {\bibinfo {author} {\bibfnamefont {H.-G.}\ \bibnamefont
  {Piao}}, \bibinfo {author} {\bibfnamefont {H.-C.}\ \bibnamefont {Choi}},
  \bibinfo {author} {\bibfnamefont {J.-H.}\ \bibnamefont {Shim}}, \bibinfo
  {author} {\bibfnamefont {D.-H.}\ \bibnamefont {Kim}}, \ and\ \bibinfo
  {author} {\bibfnamefont {C.-Y.}\ \bibnamefont {You}},\ }\href {\doibase
  10.1063/1.3658805} {\bibfield  {journal} {\bibinfo  {journal} {Appl. Phys.
  Lett.}\ }\textbf {\bibinfo {volume} {99}},\ \bibinfo {pages} {192512}
  (\bibinfo {year} {2011})}\BibitemShut {NoStop}%
\bibitem [{\citenamefont {Chaves-O'Flynn}\ \emph {et~al.}(2015)\citenamefont
  {Chaves-O'Flynn}, \citenamefont {Wolf}, \citenamefont {Pinna},\ and\
  \citenamefont {Kent}}]{Chaves:JAP_2015}%
  \BibitemOpen
  \bibfield  {author} {\bibinfo {author} {\bibfnamefont {G.~D.}\ \bibnamefont
  {Chaves-O'Flynn}}, \bibinfo {author} {\bibfnamefont {G.}~\bibnamefont
  {Wolf}}, \bibinfo {author} {\bibfnamefont {D.}~\bibnamefont {Pinna}}, \ and\
  \bibinfo {author} {\bibfnamefont {A.~D.}\ \bibnamefont {Kent}},\ }\href
  {\doibase 10.1063/1.4907241} {\bibfield  {journal} {\bibinfo  {journal} {J.
  Appl. Phys.}\ }\textbf {\bibinfo {volume} {117}},\ \bibinfo {pages} {17D705}
  (\bibinfo {year} {2015})}\BibitemShut {NoStop}%
\bibitem [{\citenamefont {Yamanouchi}\ \emph {et~al.}(2011)\citenamefont
  {Yamanouchi}, \citenamefont {Jander}, \citenamefont {Dhagat}, \citenamefont
  {Ikeda}, \citenamefont {Matsukura},\ and\ \citenamefont
  {Ohno}}]{Yamanouchi:IEEEML_2011}%
  \BibitemOpen
  \bibfield  {author} {\bibinfo {author} {\bibfnamefont {M.}~\bibnamefont
  {Yamanouchi}}, \bibinfo {author} {\bibfnamefont {A.}~\bibnamefont {Jander}},
  \bibinfo {author} {\bibfnamefont {P.}~\bibnamefont {Dhagat}}, \bibinfo
  {author} {\bibfnamefont {S.}~\bibnamefont {Ikeda}}, \bibinfo {author}
  {\bibfnamefont {F.}~\bibnamefont {Matsukura}}, \ and\ \bibinfo {author}
  {\bibfnamefont {H.}~\bibnamefont {Ohno}},\ }\href {\doibase
  10.1109/LMAG.2011.2159484} {\bibfield  {journal} {\bibinfo  {journal} {IEEE
  Magn. Lett.}\ }\textbf {\bibinfo {volume} {2}},\ \bibinfo {pages} {3000304}
  (\bibinfo {year} {2011})}\BibitemShut {NoStop}%
\bibitem [{\citenamefont {Liu}\ \emph {et~al.}(2016)\citenamefont {Liu},
  \citenamefont {Hao},\ and\ \citenamefont {Cao}}]{Liu:AIPAdv_2016}%
  \BibitemOpen
  \bibfield  {author} {\bibinfo {author} {\bibfnamefont {Y.}~\bibnamefont
  {Liu}}, \bibinfo {author} {\bibfnamefont {L.}~\bibnamefont {Hao}}, \ and\
  \bibinfo {author} {\bibfnamefont {J.}~\bibnamefont {Cao}},\ }\href {\doibase
  10.1063/1.4947132} {\bibfield  {journal} {\bibinfo  {journal} {AIP Advances}\
  }\textbf {\bibinfo {volume} {6}},\ \bibinfo {pages} {045008} (\bibinfo {year}
  {2016})}\BibitemShut {NoStop}%
\bibitem [{\citenamefont {Bogdanov}\ and\ \citenamefont
  {R\"o\ss{}ler}(2001)}]{Bogdanov:PRL_2001}%
  \BibitemOpen
  \bibfield  {author} {\bibinfo {author} {\bibfnamefont {A.~N.}\ \bibnamefont
  {Bogdanov}}\ and\ \bibinfo {author} {\bibfnamefont {U.~K.}\ \bibnamefont
  {R\"o\ss{}ler}},\ }\href {\doibase 10.1103/PhysRevLett.87.037203} {\bibfield
  {journal} {\bibinfo  {journal} {Phys. Rev. Lett.}\ }\textbf {\bibinfo
  {volume} {87}},\ \bibinfo {pages} {037203} (\bibinfo {year}
  {2001})}\BibitemShut {NoStop}%
\bibitem [{\citenamefont {Yang}\ \emph {et~al.}(2015)\citenamefont {Yang},
  \citenamefont {Thiaville}, \citenamefont {Rohart}, \citenamefont {Fert},\
  and\ \citenamefont {Chshiev}}]{Yang:PRL_2015}%
  \BibitemOpen
  \bibfield  {author} {\bibinfo {author} {\bibfnamefont {H.}~\bibnamefont
  {Yang}}, \bibinfo {author} {\bibfnamefont {A.}~\bibnamefont {Thiaville}},
  \bibinfo {author} {\bibfnamefont {S.}~\bibnamefont {Rohart}}, \bibinfo
  {author} {\bibfnamefont {A.}~\bibnamefont {Fert}}, \ and\ \bibinfo {author}
  {\bibfnamefont {M.}~\bibnamefont {Chshiev}},\ }\href {\doibase
  10.1103/PhysRevLett.115.267210} {\bibfield  {journal} {\bibinfo  {journal}
  {Phys. Rev. Lett.}\ }\textbf {\bibinfo {volume} {115}},\ \bibinfo {pages}
  {267210} (\bibinfo {year} {2015})}\BibitemShut {NoStop}%
\bibitem [{\citenamefont {Li}\ and\ \citenamefont
  {Zhang}(2004{\natexlab{a}})}]{Li:PRL_2004}%
  \BibitemOpen
  \bibfield  {author} {\bibinfo {author} {\bibfnamefont {Z.}~\bibnamefont
  {Li}}\ and\ \bibinfo {author} {\bibfnamefont {S.}~\bibnamefont {Zhang}},\
  }\href {\doibase 10.1103/PhysRevLett.92.207203} {\bibfield  {journal}
  {\bibinfo  {journal} {Phys. Rev. Lett.}\ }\textbf {\bibinfo {volume} {92}},\
  \bibinfo {pages} {207203} (\bibinfo {year} {2004}{\natexlab{a}})}\BibitemShut
  {NoStop}%
\bibitem [{\citenamefont {Li}\ and\ \citenamefont
  {Zhang}(2004{\natexlab{b}})}]{Li:PRB_2004}%
  \BibitemOpen
  \bibfield  {author} {\bibinfo {author} {\bibfnamefont {Z.}~\bibnamefont
  {Li}}\ and\ \bibinfo {author} {\bibfnamefont {S.}~\bibnamefont {Zhang}},\
  }\href {\doibase 10.1103/PhysRevB.70.024417} {\bibfield  {journal} {\bibinfo
  {journal} {Phys. Rev. B}\ }\textbf {\bibinfo {volume} {70}},\ \bibinfo
  {pages} {024417} (\bibinfo {year} {2004}{\natexlab{b}})}\BibitemShut
  {NoStop}%
\bibitem [{\citenamefont {Thiele}(1973)}]{Thiele:PRL_1973}%
  \BibitemOpen
  \bibfield  {author} {\bibinfo {author} {\bibfnamefont {A.~A.}\ \bibnamefont
  {Thiele}},\ }\href {\doibase 10.1103/PhysRevLett.30.230} {\bibfield
  {journal} {\bibinfo  {journal} {Phys. Rev. Lett.}\ }\textbf {\bibinfo
  {volume} {30}},\ \bibinfo {pages} {230} (\bibinfo {year} {1973})}\BibitemShut
  {NoStop}%
\bibitem [{\citenamefont {Song}\ \emph {et~al.}(2020)\citenamefont {Song},
  \citenamefont {Moon}, \citenamefont {Yang}, \citenamefont {Hwang},\ and\
  \citenamefont {Kim}}]{Song:APE_2020}%
  \BibitemOpen
  \bibfield  {author} {\bibinfo {author} {\bibfnamefont {M.}~\bibnamefont
  {Song}}, \bibinfo {author} {\bibfnamefont {K.-W.}\ \bibnamefont {Moon}},
  \bibinfo {author} {\bibfnamefont {S.}~\bibnamefont {Yang}}, \bibinfo {author}
  {\bibfnamefont {C.}~\bibnamefont {Hwang}}, \ and\ \bibinfo {author}
  {\bibfnamefont {K.-J.}\ \bibnamefont {Kim}},\ }\href {\doibase
  10.35848/1882-0786/ab8d0b} {\bibfield  {journal} {\bibinfo  {journal}
  {Applied Physics Express}\ }\textbf {\bibinfo {volume} {13}},\ \bibinfo
  {pages} {063002} (\bibinfo {year} {2020})}\BibitemShut {NoStop}%
\bibitem [{\citenamefont {Battiato}\ \emph {et~al.}(2010)\citenamefont
  {Battiato}, \citenamefont {Carva},\ and\ \citenamefont
  {Oppeneer}}]{Battiato:PRL_2010}%
  \BibitemOpen
  \bibfield  {author} {\bibinfo {author} {\bibfnamefont {M.}~\bibnamefont
  {Battiato}}, \bibinfo {author} {\bibfnamefont {K.}~\bibnamefont {Carva}}, \
  and\ \bibinfo {author} {\bibfnamefont {P.~M.}\ \bibnamefont {Oppeneer}},\
  }\href {\doibase 10.1103/PhysRevLett.105.027203} {\bibfield  {journal}
  {\bibinfo  {journal} {Phys. Rev. Lett.}\ }\textbf {\bibinfo {volume} {105}},\
  \bibinfo {pages} {027203} (\bibinfo {year} {2010})}\BibitemShut {NoStop}%
\bibitem [{\citenamefont {Carva}\ \emph {et~al.}(2017)\citenamefont {Carva},
  \citenamefont {Bal\v{a}\v{z}},\ and\ \citenamefont {Radu}}]{Carva:HMM_2017}%
  \BibitemOpen
  \bibfield  {author} {\bibinfo {author} {\bibfnamefont {K.}~\bibnamefont
  {Carva}}, \bibinfo {author} {\bibfnamefont {P.}~\bibnamefont
  {Bal\v{a}\v{z}}}, \ and\ \bibinfo {author} {\bibfnamefont {I.}~\bibnamefont
  {Radu}}\ }(\bibinfo  {publisher} {Elsevier},\ \bibinfo {year} {2017})\ pp.\
  \bibinfo {pages} {291--463}\BibitemShut {NoStop}%
\bibitem [{\citenamefont {Schumacher}\ \emph {et~al.}(2003)\citenamefont
  {Schumacher}, \citenamefont {Chappert}, \citenamefont {Sousa}, \citenamefont
  {Freitas}, \citenamefont {Miltat},\ and\ \citenamefont
  {Ferré}}]{Schumacher:JAP_2003}%
  \BibitemOpen
  \bibfield  {author} {\bibinfo {author} {\bibfnamefont {H.~W.}\ \bibnamefont
  {Schumacher}}, \bibinfo {author} {\bibfnamefont {C.}~\bibnamefont
  {Chappert}}, \bibinfo {author} {\bibfnamefont {R.~C.}\ \bibnamefont {Sousa}},
  \bibinfo {author} {\bibfnamefont {P.~P.}\ \bibnamefont {Freitas}}, \bibinfo
  {author} {\bibfnamefont {J.}~\bibnamefont {Miltat}}, \ and\ \bibinfo {author}
  {\bibfnamefont {J.}~\bibnamefont {Ferré}},\ }\href {\doibase
  10.1063/1.1557376} {\bibfield  {journal} {\bibinfo  {journal} {J. Appl.
  Phys.}\ }\textbf {\bibinfo {volume} {93}},\ \bibinfo {pages} {7290} (\bibinfo
  {year} {2003})}\BibitemShut {NoStop}%
\bibitem [{\citenamefont {Sampaio}\ \emph {et~al.}(2013)\citenamefont
  {Sampaio}, \citenamefont {Cros}, \citenamefont {Rohart}, \citenamefont
  {Thiaville},\ and\ \citenamefont {Fert}}]{Sampaio:NatNanotech_2013}%
  \BibitemOpen
  \bibfield  {author} {\bibinfo {author} {\bibfnamefont {J.}~\bibnamefont
  {Sampaio}}, \bibinfo {author} {\bibfnamefont {V.}~\bibnamefont {Cros}},
  \bibinfo {author} {\bibfnamefont {S.}~\bibnamefont {Rohart}}, \bibinfo
  {author} {\bibfnamefont {A.}~\bibnamefont {Thiaville}}, \ and\ \bibinfo
  {author} {\bibfnamefont {A.}~\bibnamefont {Fert}},\ }\href {\doibase
  10.1038/nnano.2013.210} {\bibfield  {journal} {\bibinfo  {journal} {Nature
  Nanotech.}\ }\textbf {\bibinfo {volume} {8}},\ \bibinfo {pages} {839}
  (\bibinfo {year} {2013})}\BibitemShut {NoStop}%
\bibitem [{\citenamefont {{Woo}}\ \emph {et~al.}(2016)\citenamefont {{Woo}},
  \citenamefont {{Litzius}}, \citenamefont {{Kr{\"u}ger}}, \citenamefont
  {{Im}}, \citenamefont {{Caretta}}, \citenamefont {{Richter}}, \citenamefont
  {{Mann}}, \citenamefont {{Krone}}, \citenamefont {{Reeve}}, \citenamefont
  {{Weigand}}, \citenamefont {{Agrawal}}, \citenamefont {{Lemesh}},
  \citenamefont {{Mawass}}, \citenamefont {{Fischer}}, \citenamefont
  {{Kl{\"a}ui}},\ and\ \citenamefont {{Beach}}}]{WOO-16}%
  \BibitemOpen
  \bibfield  {author} {\bibinfo {author} {\bibfnamefont {S.}~\bibnamefont
  {{Woo}}}, \bibinfo {author} {\bibfnamefont {K.}~\bibnamefont {{Litzius}}},
  \bibinfo {author} {\bibfnamefont {B.}~\bibnamefont {{Kr{\"u}ger}}}, \bibinfo
  {author} {\bibfnamefont {M.-Y.}\ \bibnamefont {{Im}}}, \bibinfo {author}
  {\bibfnamefont {L.}~\bibnamefont {{Caretta}}}, \bibinfo {author}
  {\bibfnamefont {K.}~\bibnamefont {{Richter}}}, \bibinfo {author}
  {\bibfnamefont {M.}~\bibnamefont {{Mann}}}, \bibinfo {author} {\bibfnamefont
  {A.}~\bibnamefont {{Krone}}}, \bibinfo {author} {\bibfnamefont {R.~M.}\
  \bibnamefont {{Reeve}}}, \bibinfo {author} {\bibfnamefont {M.}~\bibnamefont
  {{Weigand}}}, \bibinfo {author} {\bibfnamefont {P.}~\bibnamefont
  {{Agrawal}}}, \bibinfo {author} {\bibfnamefont {I.}~\bibnamefont {{Lemesh}}},
  \bibinfo {author} {\bibfnamefont {M.-A.}\ \bibnamefont {{Mawass}}}, \bibinfo
  {author} {\bibfnamefont {P.}~\bibnamefont {{Fischer}}}, \bibinfo {author}
  {\bibfnamefont {M.}~\bibnamefont {{Kl{\"a}ui}}}, \ and\ \bibinfo {author}
  {\bibfnamefont {G.~S.~D.}\ \bibnamefont {{Beach}}},\ }\href {\doibase
  10.1038/nmat4593} {\bibfield  {journal} {\bibinfo  {journal} {Nature
  Materials}\ }\textbf {\bibinfo {volume} {15}},\ \bibinfo {pages} {501}
  (\bibinfo {year} {2016})}\BibitemShut {NoStop}%
\bibitem [{\citenamefont {Finazzi}\ \emph {et~al.}(2013)\citenamefont
  {Finazzi}, \citenamefont {Savoini}, \citenamefont {Khorsand}, \citenamefont
  {Tsukamoto}, \citenamefont {Itoh}, \citenamefont {Du\`o}, \citenamefont
  {Kirilyuk}, \citenamefont {Rasing},\ and\ \citenamefont
  {Ezawa}}]{Finazzi:PRL_2013}%
  \BibitemOpen
  \bibfield  {author} {\bibinfo {author} {\bibfnamefont {M.}~\bibnamefont
  {Finazzi}}, \bibinfo {author} {\bibfnamefont {M.}~\bibnamefont {Savoini}},
  \bibinfo {author} {\bibfnamefont {A.~R.}\ \bibnamefont {Khorsand}}, \bibinfo
  {author} {\bibfnamefont {A.}~\bibnamefont {Tsukamoto}}, \bibinfo {author}
  {\bibfnamefont {A.}~\bibnamefont {Itoh}}, \bibinfo {author} {\bibfnamefont
  {L.}~\bibnamefont {Du\`o}}, \bibinfo {author} {\bibfnamefont
  {A.}~\bibnamefont {Kirilyuk}}, \bibinfo {author} {\bibfnamefont
  {T.}~\bibnamefont {Rasing}}, \ and\ \bibinfo {author} {\bibfnamefont
  {M.}~\bibnamefont {Ezawa}},\ }\href {\doibase 10.1103/PhysRevLett.110.177205}
  {\bibfield  {journal} {\bibinfo  {journal} {Phys. Rev. Lett.}\ }\textbf
  {\bibinfo {volume} {110}},\ \bibinfo {pages} {177205} (\bibinfo {year}
  {2013})}\BibitemShut {NoStop}%
\bibitem [{\citenamefont {Gerlinger}\ \emph {et~al.}(2021)\citenamefont
  {Gerlinger}, \citenamefont {Pfau}, \citenamefont {Büttner}, \citenamefont
  {Schneider}, \citenamefont {Kern}, \citenamefont {Fuchs}, \citenamefont
  {Engel}, \citenamefont {Günther}, \citenamefont {Huang}, \citenamefont
  {Lemesh}, \citenamefont {Caretta}, \citenamefont {Churikova}, \citenamefont
  {Hessing}, \citenamefont {Klose}, \citenamefont {Strüber}, \citenamefont
  {Schmising}, \citenamefont {Huang}, \citenamefont {Wittmann}, \citenamefont
  {Litzius}, \citenamefont {Metternich}, \citenamefont {Battistelli},
  \citenamefont {Bagschik}, \citenamefont {Sadovnikov}, \citenamefont {Beach},\
  and\ \citenamefont {Eisebitt}}]{Gerlinger:APL_2021}%
  \BibitemOpen
  \bibfield  {author} {\bibinfo {author} {\bibfnamefont {K.}~\bibnamefont
  {Gerlinger}}, \bibinfo {author} {\bibfnamefont {B.}~\bibnamefont {Pfau}},
  \bibinfo {author} {\bibfnamefont {F.}~\bibnamefont {Büttner}}, \bibinfo
  {author} {\bibfnamefont {M.}~\bibnamefont {Schneider}}, \bibinfo {author}
  {\bibfnamefont {L.-M.}\ \bibnamefont {Kern}}, \bibinfo {author}
  {\bibfnamefont {J.}~\bibnamefont {Fuchs}}, \bibinfo {author} {\bibfnamefont
  {D.}~\bibnamefont {Engel}}, \bibinfo {author} {\bibfnamefont {C.~M.}\
  \bibnamefont {Günther}}, \bibinfo {author} {\bibfnamefont {M.}~\bibnamefont
  {Huang}}, \bibinfo {author} {\bibfnamefont {I.}~\bibnamefont {Lemesh}},
  \bibinfo {author} {\bibfnamefont {L.}~\bibnamefont {Caretta}}, \bibinfo
  {author} {\bibfnamefont {A.}~\bibnamefont {Churikova}}, \bibinfo {author}
  {\bibfnamefont {P.}~\bibnamefont {Hessing}}, \bibinfo {author} {\bibfnamefont
  {C.}~\bibnamefont {Klose}}, \bibinfo {author} {\bibfnamefont
  {C.}~\bibnamefont {Strüber}}, \bibinfo {author} {\bibfnamefont {C.~v.~K.}\
  \bibnamefont {Schmising}}, \bibinfo {author} {\bibfnamefont {S.}~\bibnamefont
  {Huang}}, \bibinfo {author} {\bibfnamefont {A.}~\bibnamefont {Wittmann}},
  \bibinfo {author} {\bibfnamefont {K.}~\bibnamefont {Litzius}}, \bibinfo
  {author} {\bibfnamefont {D.}~\bibnamefont {Metternich}}, \bibinfo {author}
  {\bibfnamefont {R.}~\bibnamefont {Battistelli}}, \bibinfo {author}
  {\bibfnamefont {K.}~\bibnamefont {Bagschik}}, \bibinfo {author}
  {\bibfnamefont {A.}~\bibnamefont {Sadovnikov}}, \bibinfo {author}
  {\bibfnamefont {G.~S.~D.}\ \bibnamefont {Beach}}, \ and\ \bibinfo {author}
  {\bibfnamefont {S.}~\bibnamefont {Eisebitt}},\ }\href {\doibase
  10.1063/5.0046033} {\bibfield  {journal} {\bibinfo  {journal} {Appl. Phys.
  Lett.}\ }\textbf {\bibinfo {volume} {118}},\ \bibinfo {pages} {192403}
  (\bibinfo {year} {2021})}\BibitemShut {NoStop}%
\bibitem [{\citenamefont {Iwasaki}\ \emph {et~al.}(2013)\citenamefont
  {Iwasaki}, \citenamefont {Mochizuki},\ and\ \citenamefont
  {Nagaosa}}]{Iwasaki:NatComm_2013}%
  \BibitemOpen
  \bibfield  {author} {\bibinfo {author} {\bibfnamefont {J.}~\bibnamefont
  {Iwasaki}}, \bibinfo {author} {\bibfnamefont {M.}~\bibnamefont {Mochizuki}},
  \ and\ \bibinfo {author} {\bibfnamefont {N.}~\bibnamefont {Nagaosa}},\ }\href
  {\doibase 10.1038/ncomms2442} {\bibfield  {journal} {\bibinfo  {journal}
  {Nature Comm.}\ }\textbf {\bibinfo {volume} {4}},\ \bibinfo {pages} {1463}
  (\bibinfo {year} {2013})}\BibitemShut {NoStop}%
\end{thebibliography}

\end{document}